\documentclass[10pt,journal,compsoc]{IEEEtran}

\ifCLASSOPTIONcompsoc
  \usepackage[nocompress]{cite}
\else
  \usepackage{cite}
\fi

\usepackage[utf8]{inputenc}
\usepackage[printonlyused]{acronym}
\usepackage{caption}
\usepackage{multirow}
\usepackage{amsmath}
\usepackage{algorithm,algorithmicx,algpseudocode}
\usepackage{graphicx}
\usepackage{subcaption}

\usepackage{amsfonts}
\usepackage{hyperref}

\date{November 2020}

\usepackage[numbers]{natbib}

\usepackage{graphicx}
\usepackage{wrapfig}

\acrodef{3G}{Third Generation}
\acrodef{4G}{Fourth Generation}
\acrodef{5G}{Fifth Generation}
\acrodef{AI}{Artificial Intelligence}
\acrodef{API}{Application Programming Interface}
\acrodef{ARIMA}{Autoregressive Integrated Moving Average}
\acrodef{AutoML}{Automated Machine Learning}
\acrodef{BPTT}{Backpropagation Through Time}
\acrodef{BS}{Batch Size}
\acrodef{BOA}{Bayesian Optimization}
\acrodef{CI}{Confidence Interval}
\acrodef{CNN}{Convolutional Neural Networks}
\acrodef{DL}{Deep Learning}
\acrodef{GS}{Grid Search}
\acrodef{GSM}{Granger Selection Method}
\acrodef{HL}{Hidden Layer}
\acrodef{LR}{Learning Rate}
\acrodef{LSTM}{Long Short Term Memory}
\acrodef{LTE}{Long-Term Evolution}
\acrodef{MA}{Moving Average}
\acrodef{MAE}{Mean Absolute Error}
\acrodef{MBB}{Mobile Broadband Networks}
\acrodef{ML}{Machine Learning}
\acrodef{MS}{Manual Search}
\acrodef{NYU}{New York University}
\acrodef{PAM}{Partitioning Around Medoids}
\acrodef{PCA}{Principal Component Analysis}
\acrodef{RFE}{Recursive Feature Elimination}
\acrodef{RLS}{Recursive Least Squares}
\acrodef{RMSE}{Root Mean Square Error}
\acrodef{RNN}{Recurrent Neural Networks}
\acrodef{RS}{Random Search}
\acrodef{RT}{Running Time}
\acrodef{SGD}{Stochastic Gradient Descent}
\acrodef{TCP}{Transmission Control Protocol}
\acrodef{VAR}{Vector Autoregression}

\begin{document}


\title{Long Short Term Memory Networks for Bandwidth Forecasting in Mobile Broadband Networks under Mobility}

\author{Konstantinos Kousias, Apostolos Pappas, Ozgu Alay, Antonios Argyriou and Michael Riegler 
\IEEEcompsocitemizethanks{\IEEEcompsocthanksitem Konstantinos Kousias is with Simula Research Laboratory.\protect\\
Apostolos Pappas and Antonios Argyriou are with University of Thessaly.\protect\\
Ozgu Alay is with University of Oslo and Simula Metropolitan.\protect\\
Michael Riegler is with Simula Metropolitan.
}}

\IEEEtitleabstractindextext{
\begin{abstract}
  Bandwidth forecasting in Mobile Broadband (MBB) networks is a challenging task, particularly when coupled with a degree of mobility. In this work, we introduce \emph{HINDSIGHT++}, an open-source \textit{R}-based framework for bandwidth forecasting experimentation in MBB networks with Long Short Term Memory (LSTM) networks. 
  We instrument \emph{HINDSIGHT++} following an Automated Machine Learning (AutoML) paradigm to first, alleviate the burden of data preprocessing, and second, enhance performance related aspects. 
  We primarily focus on bandwidth forecasting for Fifth Generation (5G) networks. 
  In particular, we leverage \emph{5Gophers}, the first open-source attempt to measure network performance on operational 5G networks in the US.
  We further explore the LSTM performance boundaries on Fourth Generation (4G) commercial settings using \emph{NYU-METS}, an open-source dataset comprising of hundreds of bandwidth traces spanning different mobility scenarios.
  Our study aims to investigate the impact of hyperparameter optimization on achieving state-of-the-art performance and beyond.
  Results highlight its significance under 5G scenarios showing an average Mean Absolute Error (MAE) decrease of near 30\% when compared to prior state-of-the-art values.
  Due to its universal design, we argue that \emph{HINDSIGHT++} can serve as a handy software tool for a multitude of applications in other scientific fields.
\end{abstract}

\begin{IEEEkeywords}
Bandwidth Forecasting, Mobile Broadband Networks, Long Short Term Memory, Hyperparameter Optimization
\end{IEEEkeywords}}

\maketitle


\section{Introduction}
\label{sec:introduction}

\IEEEPARstart{O}{ver} the recent years, the popularity of \ac{MBB} has shown a great rise, triggering research studies and activities from both academic and industrial parties.
\ac{MBB} networks sustain a huge portion of the global wireless telecommunications in the world, including video, voice, and Internet.
Furthermore, indispensable society's operations, including healthcare, security, transport, and education, highly rely on the reliable services of \ac{MBB} networks.
They are considered to be a vital element for people that seek to stay continuously connected.

According to the Cisco Annual Internet Report\footnote{Cisco global forecast/analysis for fixed broadband, Wi-Fi, and mobile (3G, 4G, 5G technologies), $2018$-$2023$, \url{www.cisco.com/c/en/us/solutions/collateral/executive-perspectives/annual-internet-report/white-paper-c11-741490.html}}, the amount of human beings with Internet access will reach nearly two-thirds of the global population by $2023$. 
In addition, the number of global mobile devices will surpass the world’s projected population by the same year, approaching a staggering $13.1$B.
The above numbers prognosticate that the need to  effectively address tasks such as network traffic management, application provisioning, and bandwidth forecasting would greatly rise, particularly considering the occurring transition to the \ac{5G} era. 
Therefore, it is critical that mechanisms are developed to model and characterize the underlying features that define the behavior of bandwidth performance in next-generation \ac{MBB} networks from an end-to-end perspective.

The topic of bandwidth forecasting in \ac{MBB} networks has been in the spotlight for a long time.
However, it has lately become more prominent due to the fast-expanding next generation network infrastructures, which in turn, yield to a larger exchange of network data, hence, increasing the need for higher and more accurate predictions.
Furthermore, \ac{5G} introduces new frequency bands, such as the \textit{mmWave}, which drives data rates to groundbreaking limits, thus, posing new challenges to the forecasting process.
Research studies focus on integrating robust and effective algorithms, while maintaining the implementation and computational complexity in acceptable levels.
Over the years, numerous solutions have been proposed ragning from traditional statistical methods, such as the Naive, \ac{ARIMA}, and \ac{VAR}, to considerably higher complexity algorithms, namely dynamic linear models, TBATS \cite{de2011forecasting} (i.e., based on exponential smoothing), and Prophet \cite{taylor2018forecasting} (i.e., widely used by Facebook).
Within the context of \ac{AI}, both \ac{RNN} and \ac{LSTM} networks have been projected as high efficient algorithms for time series modeling.
Between the two, \ac{LSTM} networks appears to be the most fitting solution for the bandwidth forecasting study, considering the complex long term dependencies and the unforeseeable behavior that mobile traces display due to mobility.

\ac{LSTM} performance heavily relies on the selection of the model hyperparameters and the underlying dataset characteristics (i.e., periodic patterns and trends). 
Since these two aspects are tightly tied with each other, knowing how to efficiently tune an \ac{LSTM} model becomes an important but rather challenging task, thus, urging for a hyperparameter optimization solution.
Among the most popular approaches are the \ac{RS} and the \ac{BOA} \cite{bergstra2012random,pelikan1999boa}.
Furthermore, additional research studies have been published focusing on the implementation of novel hyperparameter optimization solutions \cite{domhan2015speeding,maclaurin2015gradient}.
However, despite being critical, hyperparameter optimization is frequently ignored by the majority of \ac{LSTM} related studies, since it is considered a time consuming process.

In this work, we present \emph{HINDSIGHT++}, a novel, light-weight, versatile, and open-source \textit{R}-based framework for bandwidth forecasting experimentation in \ac{MBB} networks with \ac{LSTM} networks\footnote{\emph{HINDSIGHT++} inherits  core analytical concepts (\ac{LSTM} design, hyperparameter optimization) from the \emph{HINDSIGHT} framework, \url{www.bitbucket.org/konstantinoskousias/hindsight} \cite{kousias2018hindsight}. However, it stands on its own since its main focus is to provide insights on the performance of \ac{LSTM} networks for bandwidth-related forecasting tasks in next-generation \ac{MBB} networks.}. 
The main source of motivation behind the design of  \emph{HINDSIGHT++} is two-fold.
First, to limit the implementation complexity of \ac{LSTM} networks, and second, to minimize the predictive error.
Toward this goal, we adopt the concept of \ac{AutoML}, an acronym used to describe the process of data pipeline automation.
In particular, \ac{AutoML} covers the complete learning routine, from raw data preprocessing to the final model production.
Moreover, \emph{HINDSIGHT++} is compliant with parallel and multivariate data and supports a wide forecasting horizon. 
Last, it offers different hyperparameter optimization options and \ac{LSTM} variants.
The source code of \emph{HINDSIGHT++} is structured in a dynamic fashion, so it can accommodate new features and algorithms. 
Users are encouraged to integrate more libraries and software components to satisfy their cause, but also to enrich the capabilities of the framework.

We study the topic of bandwidth forecasting in commercial \ac{5G} networks under mobility, while we revisit \ac{4G} scenarios and draw additional insights by performing a comparative analysis between the two technology standards. 
Thereupon, we leverage two open-source datasets, \emph{5Gophers}, which features network measurements from operational \ac{5G} networks in the US, and NYU Metropolitan Mobile Bandwidth Trace (\emph{NYU-METS}), a counterpart dataset that comprises of hundreds of \ac{4G} bandwidth measurements in the wild. 
We adopt systematic investigation to showcase the significance of hyperparameter optimization at each use case when compared to existing state-of-the-art approaches. 

The main contributions of this work are:
\begin{itemize}
    \item We design and implement \emph{HINDSIGHT++}, a light-weight LSTM based framework for bandwidth forecasting in \ac{MBB} networks.
    \item We carry out a large-scale study of bandwidth forecasting in \ac{MBB} networks using \emph{HINDSIGHT++}. We focus on different mobility scenarios and perform a comparative analysis between real-world operational \ac{4G} and \ac{5G} networks.
    \item We investigate the potential of hyperparameter optimization and explore the performance boundaries of different \ac{LSTM} variants attmepting to further improve their performance in a \ac{MBB} environment.
    \item Last, we open-source \emph{HINDSIGHT++} to the community allowing for further experimentation with a variety of datasets and applications \cite{repo}.
\end{itemize}

The rest of the paper is organized as follows.
Section \ref{sec:background} provides background information on \ac{LSTM} networks and summarizes related literature on network bandwidth forecasting and \ac{LSTM} applications. 
Section \ref{sec:design} outlines the design of \emph{HINDSIGHT++}, while Section \ref{sec:validation} provides an overview of the software implementation and validation details. 
Performance evaluation takes place in Section \ref{sec:performance}, while the study is concluded in Section \ref{sec:conclusions}.

\section{Background and Related Work}
\label{sec:background}

In this section, we summarize literature work concerning bandwidth and performance forecasting tasks in mobile networks.
In addition, we provide a brief background synopsis on \ac{LSTM} networks and showcase how they are used within the context of networking.

\subsection{Related Work on Performance Forecasting in Mobile Networks} 

Over the years, several studies have focused on modeling the performance of mobile networks.
Even though most of them tackle the problem from slightly separate angles, they share a common objective, i.e., to minimize the error function and provide insights on the complexity of the mobile ecosystem.

A supervised \ac{ML} solution for downlink throughput prediction in \ac{MBB} networks is proposed in \cite{Kousias2019}. 
The authors argue that \ac{ML} can be used as a tool for significantly reducing the data volume consumption over the network, while maintaining the predictive error in acceptable levels. 
Furthermore, in \cite{raida2018constant}, Raida et. al  leverage constant rate probing packets to estimate available bandwidth in a controlled \ac{LTE} environment, while in \cite{raida2019constant}, they extend their work by further testing and validating the developed framework in live \ac{LTE} networks.
In \cite{Maier2019}, Maier et al. introduce a novel \ac{AI} model with feed-forward neural networks for both  downlink and uplink bandwidth forecasting.
To find a fitting speed test duration, the authors investigate two scenarios.
First, they train a model with a duration fixed to a value lower than the default, while second, they dynamically determine a duration by using the results of a pretrained neural network model. 
Experimental approaches studying the problem of bandwidth prediction have also been addressed in  \cite{liu2015empirical, mirza2007machine, samba2017instantaneous, wei2018trust, riihijarvi2018machine, linder2016using, rattaro2010throughput}.
Beyond empirical-based studies, theoretical models have also been published.
In \cite{gao2018prophet}, Gao et al. introduce a theoretical learning based throughput prediction system for reactive flows, while authors in \cite{bui2014model} propose a novel stochastic model for user throughput prediction in \ac{MBB} networks that considers fast fading and user location. 

\subsection{Background on \ac{LSTM} Networks}

The inception of \acp{RNN} was triggered by the ever-increasing need to effectively tackle time series forecasting applications with novel \ac{AI} algorithms.
Their central mechanism falls heir to artificial neural networks, a sophisticated and rather complex ecosystem composed by a collection of connected nodes, known as \textit{artificial neurons}, that was inspired by closely observing the core operations of the human brain. 
\acp{RNN} introduced revolutionary feedback loops that would allow information to persist or vanish as it flowed from one network to another.
The innovative and unique architectural design of \acp{RNN} was embraced by communities in several fields spanning different applications, such as image captioning, speech and text recognition, and led to immensely increasing their popularity \cite{graves2013speech, gregor2015draw,mikolov2010recurrent,du2015hierarchical}.

Despite the massive success and hype around \acp{RNN} though, it was not long after Bengio et al. addressed a critical flaw affecting tasks with long range dependencies \cite{bengio1994learning}. 
In particular, by using both theoretical and empirical evidence, the authors showed that \acp{RNN} and gradient based algorithms lack the ability to learn sequences spanning long-intervals.
The gradient descent shortcoming, also known as the problem of vanishing gradients, was also re-addressed seven years later by Hochreiter, Bengio et al. \cite{hochreiter2001gradient}. 
This finding launched research initiatives that led to the introduction of \ac{LSTM} networks in $1997$. 

\begin{figure}[!htb]
  \centering
    \includegraphics[keepaspectratio,width = 0.8\linewidth]{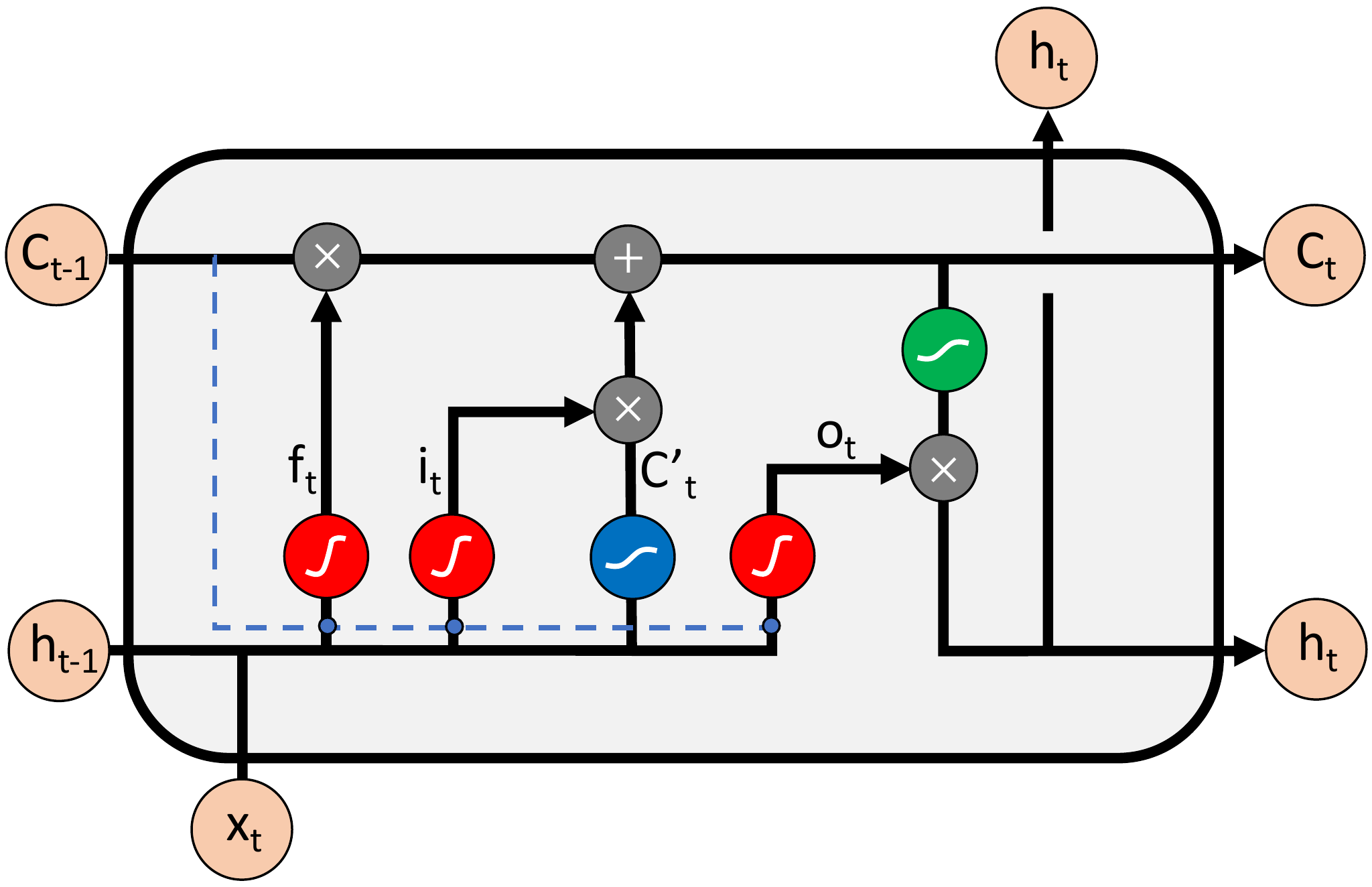} 
      \caption{High-level representation of an \ac{LSTM} memory block. Peephole connections are rendered by the blue dotted lines.}
  \label{figure:cell}
\end{figure}

While \ac{LSTM} networks are grouped under the big umbrella of \acp{RNN}, thus inheriting most of the latter's key mechanisms, they introduce additional novel components explicitly added for capturing long term dependencies.
The prime difference between the two algorithms is the number of neural network layers. 
The structure of a traditional \ac{RNN} architecture is fairly simplistic and features a single neural network layer (i.e., tangent function).
On the contrary, an \ac{LSTM} architecture inserts additional complexity by engaging a total number of four neural networks (i.e., one tangent and three sigmoid functions) alongside a collection of interconnecting elements.
The above components interact with each other in a unique manner resulting in a perplexed ecosystem that is known as \textit{\ac{LSTM} cell}.
The main objective of an \ac{LSTM} cell is to convey or remove information that flows along the neighbouring cells.
Since \ac{LSTM} cells establish a memory-wise mechanism, they are also referred in the literature as \textit{memory blocks}.
Figure \ref{figure:cell} shows a schematic of an \ac{LSTM} memory block.

The three central components responsible for exploiting the memory aspect of an \ac{LSTM} cell are the \textit{forget}, \textit{input}, and \textit{output} gates. 
As its name suggests, the forget gate is responsible for removing redundant information from the cell. 
Research studies have shown that alongside the output activation function, forget gate constitutes the most critical element of the cell \cite{greff2016lstm}.
Contrarily, the input gate acts as a regulator mechanism whose task is to preserve all vital information.
Last, the output gate is used to decide the cell’s final state by using similar functionality as the forget gate. 
In other words, the output gate is the link between two consecutive \ac{LSTM} memory blocks.

\subsection{Network Performance Forecasting with \ac{LSTM} Networks} 
Over the past years \ac{LSTM} networks have been successfully applied for network performance related tasks.
Cui et al. proposed a stacked bidirectional \ac{LSTM} architecture for network-wide traffic estimation \cite{cui2018deep}, while authors in \cite{mei2019realtime} studied the applicability of \ac{LSTM} for a real-time bandwidth prediction problem.
Furthermore, Zhao et. al. focused on short-term traffic estimation by considering temporal–spatial correlation \cite{zhao2017lstm}, whereas, authors in \cite{wang2017spatiotemporal} proposed a hybrid framework that combines \ac{LSTM} networks and an auto-encoder based model for network prediction from a spatio-temporal angle.
Additional work on mobile traffic forecasting has been further proposed in \cite{ma2015long,yu2017deep,fu2016using}.

Distinct from the preceding studies, this paper stands out and distinguishes itself by putting focus on the hyperparameter optimization aspect applied on data from different countries, network operators, technology standards, and mobility scenarios.
In addition, to the best of our knowledge, this is the first study that attempts to perform a  bandwidth forecasting comparative analysis between \ac{4G} and \ac{5G} networks under mobility with \ac{LSTM} networks. 
Last, we make available to the community an open-source framework which follows an end-to-end \ac{AutoML} paradigm and allows for ample experimentation with \ac{LSTM} networks.


\section{Framework Design}
\label{sec:design}

\begin{figure*}[!htb]
  \centering
    \includegraphics[keepaspectratio,width = 0.8\linewidth]{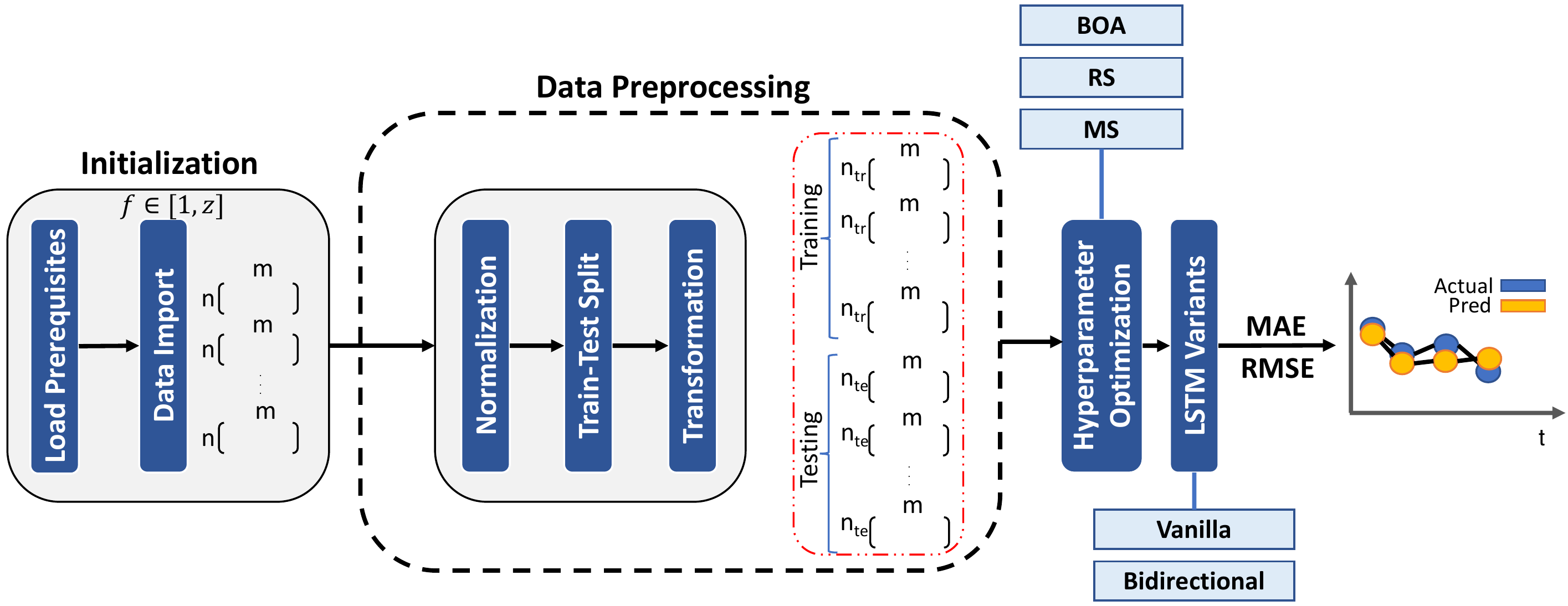} 
      \caption{A high-level representation of the \emph{HINDSIGHT++} control flow diagram. The operations for each block are described in Section \ref{sec:design}, highlighted with a matching title. Dashed lines delimit the data preprocessing stage.}
  \label{figure:flow}
\end{figure*}

This section's intent is to provide an overview of \emph{HINDSIGHT++}.
Figure \ref{figure:flow} is a high-level control flow diagram that displays the interconnection of the different blocks, alongside complementary information, including matrix and parameter notation.
We encourage the reader to use this diagram as a guideline or reference point when navigating through the current section.

\emph{HINDSIGHT++} is designed in a universal fashion so it supports both short and long range forecasts.
As a result, we adopt an \textit{encoder-decoder} scheme featuring two \ac{LSTM} architectures that operate as a pair.
The \textit{encoder-decoder} design was originally introduced to address sequence to sequence (known as \textit{seq2seq}) learning problems \cite{cho2014learning}. 
In a nutshell, \textit{encoder's} objective is to read and translate the input sequence into a sequence with a fixed length destined to be used as input to the \textit{decoder}, which in turn, it is responsible for generating an output value for each of the prediction steps. 
The number of \acp{HL} and neurons across the two blocks can differ while additional \ac{LSTM} variants may be used (e.g., Bidirectional \ac{LSTM} networks).


\subsection{Initialization}
\label{sec:initialization}

The objective of this first phase is two-fold.
First, to provide support with all required software components, and second, to initiate the data import process.

\subsubsection{Load Prerequisites}
\emph{HINDSIGHT++} relies on a number of open-source packages, or else, a collection of functions, compiled code, and data, available in \textit{CRAN}. 
To guarantee software stability and robustness, it is critical that all dependencies are pre-installed in the system.
Therefore, we enable a verification process that automatically locates, installs, and loads any missing packages and libraries.

\subsubsection{Data Import}
Next is the data import process where users select a number of $f\in[1,z]$ \textit{.csv} files via an interactive window. 
Due to its universal design, \emph{HINDSIGHT++} operates under the assumption that the selected datasets are parallel\footnote{In this work, we do not cover the parallel aspect of the framework since it is not relevant for the bandwidth prediction problem.}.
In addition, equidimensional datasets are required, since \ac{LSTM} models expect input arrays with equal sizes. 
The last prerequisite is that the dependent variable  locates in the first array position followed by any number of additional of regressors. 
We annotate the matrix dimensions for each dataset as $d_{f}\in{\mathbb{R}^{n \times m}}$, where $n$ and $m$ represent the number of samples and features, respectively. 

\subsection{Data Preprocessing}
\label{sec:preprocessing}

The process of data preparation in \ac{LSTM} networks is rather perplexed time consuming.  
The complexity factor significantly increases considering the different data format combinations (e.g., multivariate, parallel) and parameters (e.g., lags, prediction steps). 
Figure \ref{figure:flow} (dashed frame) encloses all steps followed for providing compatibility with the \ac{LSTM} models. 
In line with the \ac{AutoML} concept, data preprocessing is treated as a black box\footnote{In addition, \emph{HINDSIGHT++} supports \textit{one-hot-encoding} for converting categorical variables to binary features and a number of options for dealing with missing data (e.g., interpolation, \ac{MA}, Kalman filtering, etc.). We do not cover these additional aspects in detail, since they are not part of our data.}.



\subsubsection{Normalization}

Multivariate data analysis often involves features with  variable scales.
Since \ac{LSTM} weight allocation is prone to such data,  it is critical that we consider a transformation function leveraging one of the following state-of-the-art normalization methods, known as \textit{min-max}, \textit{z-score}, and \textit{tanh} \cite{jain2005score}.

\subsubsection{Train-Test split}

The next phase involves splitting each dataset into a training ($80\%$) and a testing ($20\%$) set.
Both of these percentages can be configured as appropriate.
We define the dimensions for each set as $d^{tr}_{f}\in{\mathbb{R}^{n_{tr} \times m}}$ and $d^{te}_{f}\in{\mathbb{R}^{n_{te} \times m}}$, where $n_{tr}$ and $n_{te}$ are the number of samples for the training and testing set, respectively, while $m$ represents the number of features.

\begin{table}[!htb]
\begin{tabular}{l|l|l}
&$2$\textbf{-D matrices}&$3$\textbf{-D matrices}\\
\hline\hline
\multirow{2}{*}{Training}&$X_{TR}\in{\mathbb{R}^{n_{TR} \times nlags*m*z}}$&$X_{TR'}\in{\mathbb{R}^{n_{TR} \times nlags \times m*z}}$\\ 
&$Y_{TR}\in{\mathbb{R}^{n_{TR} \times msteps*z}}$&$Y_{TR'}\in{\mathbb{R}^{n_{TR} \times msteps \times z}}$\\
\multirow{2}{*}{Testing}&$X_{TE}\in{\mathbb{R}^{n_{TE} \times nlags*m*z}}$ &  $X_{TE'}\in{\mathbb{R}^{n_{TE} \times nlags \times m*z}}$\\ 
&$Y_{TE}\in{\mathbb{R}^{n_{TE} \times msteps*z}}$ &  $Y_{TE'}\in{\mathbb{R}^{n_{TE} \times msteps \times z}}$
\end{tabular}
\caption{Summary of the $2$-D and $3$-D matrix dimensions, where $n_{TR} = n_{tr} - (nlags + msteps - 1)$ and $n_{TE} = n_{te} - (nlags + msteps - 1)$.}
\label{table:dim}
\end{table}

\subsubsection{Transformation}

Finally, data undergoes transformation and reshaping functions to comply with the \ac{LSTM} input requirements.
The outcome of this process are the $2$-D and $3$-D matrices summarized in Table \ref{table:dim}.
$X$ and $Y$ represent the \ac{LSTM} input and output, respectively.

\subsection{Hyperparameter Optimization}
\label{sec:hyperparameter}

Even though hyperparameter optimization is a rather time consuming process, it can be key for achieving state-of-the-art performance and beyond.
As human beings, we have a hard time to grasp, handle, and visualize multi-dimensional spaces, hence, manual hyperparameter tuning is often a rugged assignment. 
Notable hyperparameters examples include the \acp{HL}, neurons, \ac{BS}, and \ac{LR}. 

To the best of our knowledge, little research has been conducted to provide guidelines on the  \ac{LSTM} hyperparameter selection, since most studies focus on particular architectures and datasets \cite{greff2016lstm}. 
Therefore, we are moving toward iterative based methods in an effort to improve the \ac{LSTM} performance.
In the following, we highlight the most popular hyperparameter optimization approaches. 

\begin{itemize}
    \item \textbf{\ac{MS}}. Despite not being an optimization algorithm per definition, manual hyperparameter tuning is still possible in instances where prior knowledge about the system under study exists. 
    \ac{MS} is the least efficient option from the list since it is quite occasional and it usually requires extensive \textit{trial and error}. 
    On the positives, it is a very simplistic approach and involves minimum coding effort.
    \item \textbf{\ac{GS}}. The \textit{brute force} concept refers to the practice of trying out all possible solutions on the way to the global optimum.  
    The equivalent algorithmic method within the context of hyperparameter optimization is known as \ac{GS}.
    On the one hand, \ac{GS} outperforms every optimization method in the market since it always tracks down the best solution. 
    On the other hand, as the number of hyperparameters increase, it suffers from severe scalability issues.
    Additionally, it requires considerable computational resources, that even with a powerful computer, optimization can take days or even weeks.
    Due to the above, \ac{GS} is not currently supported.
    \item \textbf{Random Search (RS)}. Instead of evaluating all possible hyperparameter combinations, \ac{RS} iterates over a smaller sample  \cite{bergstra2012random}. 
    As the number of iterations increase, the probability of converging to a better solution increases as well. 
    The number of iterations required to reach to a good solution depends on several factors including the size of the hyperparameter space, search range, data complexity, and so forth. 
    \ac{RS} has a fairly simple implementation and does not require any scientific knowledge for the hyperparameters under study. 
    In addition, it provides significant gains in terms of computational complexity, since it only tests a limited number of combinations, which is a safe approach under the assumption that non all hyperparameters are equally important.
    As a rule of thumb, there is a $95\%$ chance that \ac{RS} will reach to a good solution with only $60$ iterations.
    \item \textbf{Bayesian Optimization (BOA).} A different alternative for complex and noisy objective functions is the \ac{BOA} algorithm, a probabilistic approach that incorporates learning for hyperparameter optimization \cite{pelikan1999boa}. 
    \ac{BOA} has its roots on the Bayes theorem which exploits the concept of conditional probability for making new predictions. 
    Akin to \ac{RS}, a number of iterations  is required for converging to a good solution. 
    Over the years, the popularity of \ac{BOA} has increased, finding application to a wide range of problems including robotics, interactive animation, and \ac{DL} related tasks \cite{snoek2012practical}.
\end{itemize}

Table \ref{table:hyperparameters} summarizes the available hyperparameters alongside their search range and a short description, while Table \ref{table:parameters} references all remaining framework parameters.

\begin{table}[!htb]
\begin{tabular}{|l|l|l|l|l|}
\hline
\textbf{ID}&\textbf{Name}&\textbf{Short Description}&\textbf{Range}&\textbf{Init}\\
\hline
\hline
$1$&\texttt{units1}&No. neurons (\ac{HL}-$1$)&$16-1024$&$256$\\
$2$&\texttt{units2}&No. neurons (\ac{HL}-$2$)&$16-1024$&$128$\\
$3$&\texttt{units3}&No. neurons (\ac{HL}-$3$)&$16-1024$&$-$\\
$4$&\texttt{lr}&\ac{LR}&$0.00001-0.01$&$0.001$\\ 
$5$&\texttt{nepochs}&No. epochs&$50-100$&$50$\\
$6$&\texttt{bs}&\ac{BS}&$8-128$&$8$\\
$7$&\texttt{nlayers}&No. \acp{HL}&$1-3$&$2$\\
\hline
\end{tabular}
\caption{\small List of the available hyperparameters alongside a short description and their default search range. \textit{Init} represents the hyperparameter values reported in \cite{mei2019realtime}.}
\label{table:hyperparameters}
\end{table}

\begin{table}[!htb]
\begin{tabular}{|l|l|l|l|}
\hline
\textbf{ID}&\textbf{Name}&\textbf{Short Description}&\textbf{Init}\\
\hline \hline
$1$&\texttt{nfeatures}&No. features& $1$\\
$2$&\texttt{nlags}&No. lags&$5$\\
$3$&\texttt{msteps}&No. prediction steps&$1$\\
$4$&\texttt{norm}&Normalization&\textit{min-max}\\
$5$&\texttt{network}&\ac{LSTM} architecture&\textit{Vanilla}\\
$6$&\texttt{act}&Activation function&\textit{tanh}\\
$7$&\texttt{act}&Optimization algorithm&\textit{Adam}\\
$8$&\texttt{split}&Train-Test split&$0.8$ \\
$9$&\texttt{valsplit}&Validation split&$0.2$\\
$10$&\texttt{hyper}&Hyperparameter optimization&\ac{RS}\\
$11$&\texttt{niter}&No. iterations&$50$\\
\hline
\end{tabular}
\caption{\small List of \emph{HINDSIGHT++} parameters alongside a short description and their initial values.}
\label{table:parameters}
\end{table}

\subsection{\ac{LSTM} Variants}
\label{sec:variants}

Over the years, a number of \ac{LSTM} variants have been proposed targeting to further improve the performance of particular applications.
Besides the Vanilla \ac{LSTM} implementation, \emph{HINDISIGHT++} enables the experimentation with Bidirectional \ac{LSTM}, a variant that enables a duplicate \ac{LSTM} sequence for traversing the network using opposite directions.
In the following, we provide a short reference summary for each of the two variants.

\begin{itemize}
    \item \textbf{Vanilla \ac{LSTM}:}
    A few years after the inception of \ac{LSTM} networks back in $1997$, the memory blocks underwent slight modifications.
    Within a two-year period,  Gers et. al. proposed the integration of forget gate and peephole connections \cite{gers1999learning,gers2000recurrent}.
    In addition, in \cite{graves2005framewise}, Graves et al. presented the full \ac{BPTT} training which finalized the Vanilla \ac{LSTM} design. 
    \item \textbf{Bidirectional \ac{LSTM}:} Following the concept of Bidirectional \acp{RNN}, Bidirectional \ac{LSTM} networks introduce an additional hidden layer to process each sequence both in a forward and a backward direction \cite{schuster1997bidirectional}, ultimately training two sequences at once.
    Bidirectional \ac{LSTM} networks are known to outperform the Vanilla counterpart networks in tasks such as phoneme, speech and audio recognition \cite{graves2005framewise,graves2013hybrid,marchi2014multi}, while, they were also recently proposed for network traffic forecasting \cite{cui2018deep}.
    
\end{itemize}


\section{Framework Validation}
\label{sec:validation}

We organize the following content into three main parts.
First, we provide an overview of \emph{HINDSIGHT++} technical and implementation details. 
Next, we discuss the training, validation, and testing process, while we outline the key \ac{LSTM} parameters.
Last, we present the available benchmarking options alongside a brief visualization description.

\subsection{Implementation and technical concerns}
The code of \emph{HINDSIGHT++} is exclusively written in \emph{R} and a proof of concept implementation is available in \cite{repo}.
The backbone is the \textit{CRAN} interface to \texttt{Keras}, a high-level, user-friendly \ac{API} which enables quick and dynamic experimentation with \ac{DL} algorithms, while it further allows for easy neural network architectural prototyping.
\texttt{Keras} is designed to run on top of multiple back-end environments, including \texttt{Theano}, \texttt{CNTK}, and \texttt{Tensorflow} (default). 
Furthermore, it offers the option for executing the code either on top of CPUs or GPUs.

\subsection{Learning Process}
Training of neural networks is an iterative process that involves identifying unique data properties and learning the model parameters.
At each iteration, we track both training and validation error, which we use as bias and variance measures, or else, indicators of underfitting or overfitting.
The validation set always reflects the last portion of the initial training set. 
We test the final model on an never-seen-before testing set and use the \ac{MAE} and \ac{RMSE} as our error metrics.

In the following, we enlist the core training parameters.

\begin{itemize}
\item \textbf{Activation functions}: \emph{HINDSIGHT++} supports four activation functions, i.e.,  \textit{tanh},
\textit{relu}, known for its fast convergence \cite{krizhevsky2012imagenet}, \textit{sigmoid}, and \textit{softmax}, primarily used for classification related tasks.
\item \textbf{\acp{HL}}: We limit the number of \acp{HL} between zero and three. 
Research studies have shown that they are sufficient for achieving top-level performance in the majority of tasks. 
\item \textbf{Epochs}: The number of epochs define the number of forward and backward passes during the training process. The default search range for epochs ranges between $50$ and $100$.
\item \textbf{\ac{BS}}: The term \ac{BS} refers to the number of samples to be propagated in the network. Lower values significantly increase the \ac{RT} complexity, therefore, we restrain it between $8$ and $128$. 
\item \textbf{\ac{LR}}: Last, the \ac{LR} determines to what extent newly acquired information overrides old information \cite{keskar2016large,smith2017don}. Its search range is between $0.00001$ and $0.01$.
\end{itemize}

In addition, we use the \textit{Adam} gradient descent optimization algorithm throughout all our experiments. 
Hovewer, users can experiment with other options, including \textit{\ac{SGD}}, \textit{Adagrad}, and \textit{RmsProp}.

\subsection{Benchmarking and Visualization}

We enable dedicated timers to allow tracking the end-to-end \ac{RT} complexity with high precision.
Such a feature is particularly critical during large-scale studies where the hyperparameter search space can be large.
In addition, timers can be used for measuring the impact of \acp{HL} or neurons during training, GPU/CPU performance benchmarking, and many more.
We further provide an automated solution for storing the final models in an \textit{.hd5f} format. 

Time series visualization can be used to highlight certain phenomena at glance, including periodic patterns or trends, extreme outliers, odd observations, abrupt changes over time, and so forth.
Moreover, it allows for visual inspection of the developed models to assess how well they fit the datasets under study. 
A different color scheme is used to distinguish  the training, validation, and testing set.  

\section{Performance Evaluation}
\label{sec:performance}

At present time, \ac{4G} networks are considered the norm of  wireless telecommunication systems sustaining a big portion of society's global network functions.
However, the recent advances in technology indicate that in the upcoming decade \ac{5G} will become the next big \textit{thing}  with several network operators already moving toward its adoption.
Therefore, it is dire that we identify the principal network characteristics and trends between the two technology standards and show to what extent they dictate the performance of \ac{LSTM} networks.
In addition, our experiments aim to identify the potential gains of hyperparameter optimization in delivering more efficient and robust models. 
Next, we provide a description of the two datasets under study.

\subsection{Datasets}

We exploit \emph{$5$Gophers}\footnote{\url{https://fivegophers.umn.edu/www20}} \cite{narayanan2020first,narayanan2020lumos5g}, the first open-source dataset that attempts to study the performance of \ac{5G} in commercial settings, featuring three operational operators (two mmWave and one mid-band carrier) in the US.
\emph{$5$Gophers} allows for a wide range of analyses, including impact of mobility, handoffs, network operator performance, and many more.  
In this study, we focus on a single \textit{walking}, and two \textit{driving} traces (i.e., medium mobility between $20$ to $50$Km), since we are primarily interested in the mobility aspect.
The \textit{walking} trace comprises from both \ac{4G} and \ac{5G} readings, followed by additional features, such as cell and handover identifiers\footnote{We conducted an offline  multivariate campaign to investigate whether the additional features have a positive impact in  performance. Results revealed that the error improvement was insignificant. In addition, as expected, we observed a slight increase in the \ac{RT} complexity.}. 
The experimental setup under medium mobility features three \textit{SGS$10$} devices mounted on a vehicle's front windshield and measuring the iPerf performance for three major operators in Atlanta, i.e., Sprint, T-Mobile, and Verizon. 
A bandwidth estimation measure is recorded every $120$sec.
All traces alongside the number of available samples are reported in Table \ref{table:5Gophers}.

\begin{table}[!htb]
\begin{tabular}{|l|l|l|l|l|}
\hline
& \multicolumn{3}{|c|}{\ac{RT}} & \\
\hline
\textit{Trace} & \textit{$RS_{VL}$} & \textit{$RS_{BD}$} & \textit{$BOA_{VL}$} & \textit{samples}\\
\hline\hline 
\textbf{SprintA} & $9$ & $16$ & $52$ & $120$\\
\textbf{SprintB} & $8$ & $19$ & $45$ & $120$\\
\textbf{VerizonA} & $10$ & $16$ & $49$ & $120$\\
\textbf{VerizonB} & $16$ & $19$ & $49$ & $120$\\
\textbf{Walking} & $37$ & $29$ & $58$ & $603$\\
\hline
\end{tabular}
\caption{\emph{5Gophers}. i. Per trace \ac{RT} complexity [in min] along the \ac{LSTM} configurations under study, and ii. number of available samples.}
\label{table:5Gophers}
\end{table}

We complement our study with \emph{NYU-METS} \cite{NYU-METS}, an open-source dataset that comprises of \ac{4G} \ac{LTE} bandwidth measurements carried out in the \ac{NYU} metropolitan area. 
\emph{NYU-METS} covers several transportation modes including bus, subway, and ferry. 
The experimental setup features an \ac{LTE}-enabled mobile device and a remote server located at the \ac{NYU} lab.
\ac{TCP} measurements are carried out using the \textit{iPerf} cross-platform tool with a sampling rate of one second. 
Likewise, Table \ref{table:NYU-METS} lists the available bandwidth traces \cite{NYU-METS}. 
In \cite{mei2019realtime}, Mei et al. introduce \emph{NYU-METS} and conduct an \ac{LSTM} related study for realtime bandwidth prediction in \ac{4G} networks.
The authors exploit a simplified \ac{LSTM} model that features two \acp{HL} for modeling the temporal behavior in each scenario. 
Complementary analysis showcasing the challenges of multi-step prediction is also presented.
Results show that \ac{LSTM} networks outperform state-of-the-art prediction algorithms including \ac{RLS} and harmonic mean. 

\begin{table}[!htb]
\begin{tabular}{|l|l|l|l|l|}
\hline
& \multicolumn{3}{|c|}{\ac{RT} (min)} & \\
\hline
\textit{Trace} & \textit{$RS_{VL}$} & \textit{$RS_{BD}$} & \textit{$BOA_{VL}$} & \textit{samples}\\
\hline\hline 
\textbf{7TrainA} & $66$ & $136$ & $110$ & $4910$\\
\textbf{7TrainB} & $50$ & $84$ & $69$ & $3316$\\
\textbf{Bus57} & $38$ & $83$ & $77$ & $2403$\\
\textbf{Bus62A} & $39$ & $74$ & $82$ & $3293$\\
\textbf{Bus62B} & $28$ & $45$ & $66$ & $1452$\\
\textbf{NTrain} & $44$ & $68$ & $95$ & $2404$\\
\textbf{LI} & $28$ & $52$ & $72$ & $1447$\\
\hline
\end{tabular}
\caption{\emph{NYU-METS} i. Per trace \ac{RT} complexity [in min] along the \ac{LSTM} configurations under study, and ii. number of available samples.}
\label{table:NYU-METS}
\end{table}

\textbf{Experimental Design:} All experiments are carried out on an \textit{x86-64} architecture with a $16$GB RAM while an \textit{NVidia Titan X} graphics card is used to accelerate data preprocessing.
The operating system is based on a Linux Ubuntu 16.04.6 \textit{LTS} distribution.
All reported \ac{RT} values could significantly vary under different architectural designs.

\subsection{Experimental setup}

We outline the experimental setup which we divide into two parts.
The first part establishes the baseline \ac{LSTM} performance, while the second part investigates the performance gains of hyperparameter optimization.

\textbf{Setting the Baseline:} For the baseline experiments ($MS_{VL}$), we adopt a stacked Vanilla architecture with two \acp{HL} and the model hyperparameters\footnote{The number of epochs alongside \ac{BS} are not reported in the paper, therefore, we set them to a predefined value.} reported in \cite{mei2019realtime} and illustrated in Table \ref{table:hyperparameters}. Note that these hyperparameters was prior optimized (e.g., by means of sensitivity analysis or by using indicative values based on previous experience with datasets of similar nature) on the same dataset \cite{mei2019realtime}. 
\ac{LSTM} experimentation takes place in a \textit{per-trace} fashion, i.e., the first $x\%$ part of the sequence is used for training and validation, while the remaining $y\%$ is used for testing.
In addition, we adopt a walk-forward validation, or else, rolling forecast approach, where each prediction value is used as new input to the model for forecasting the next step. 
Last, we overcome the random \ac{LSTM} weights initialization bias by repeating each baseline experiment $25$ times and reporting the median values for both of our error metrics\footnote{Statistical confidence can be strengthened by increasing the number of iterations. However, this would significantly add to the \ac{RT} overhead.}.

\textbf{Toward Performance Optimization:} We repeat the experimental campaign by enabling hyperparameter optimization while testing the available \ac{LSTM} variants.
Our goal is to test whether the hypothesis stating that a set of fixed model hyperparameters is sub-optimal for modelling trends from different technology standards, mobility scenarios, and network operators holds.
We annotate each configuration as $H_{A}$, where $H\in[MS,RS,BOA]$ represents the hyperparameter optimization algorithm and $A\in[VL,BD]$ denotes the \ac{LSTM} variant.
The number of iterations for both \ac{RS} and \ac{BOA} is set to $50$, while we decide the final model using the \ac{MAE} validation score.

\subsection{Performance Analysis}

\begin{figure*}[!htb]
  \centering
    \includegraphics[keepaspectratio,width = 0.6\linewidth]{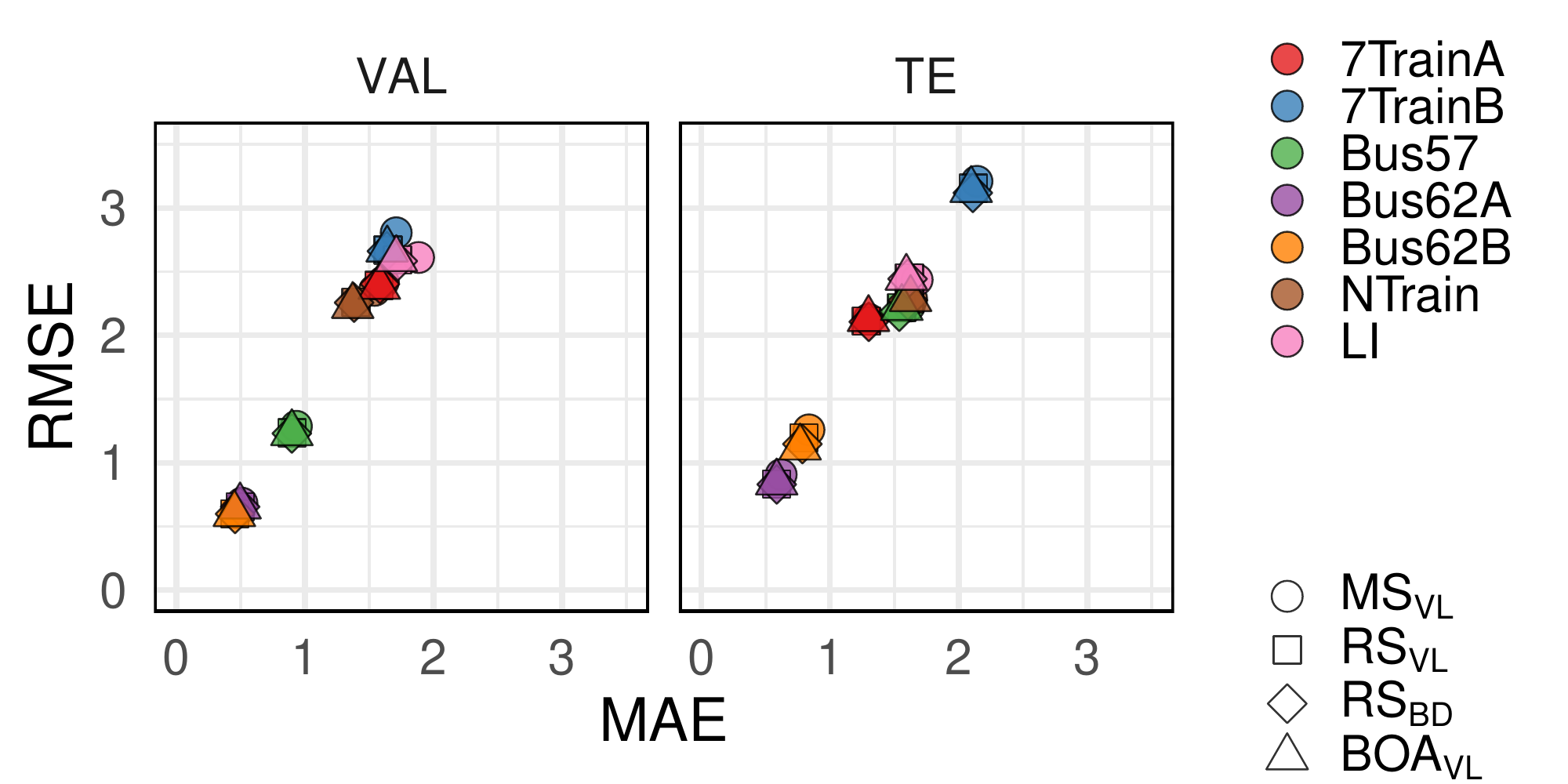} 
      \caption{\emph{NYU-METS}. \ac{MAE} versus \ac{RMSE} colored by trace and grouped per configuration. Each subplot maps to the validation (\textit{VAL}) and testing (\textit{TE}) error, respectively.}
  \label{fig:NYU-METS_error}
\end{figure*}

\begin{figure*}[!htb]
  \centering
    \includegraphics[keepaspectratio,width = 0.65\linewidth]{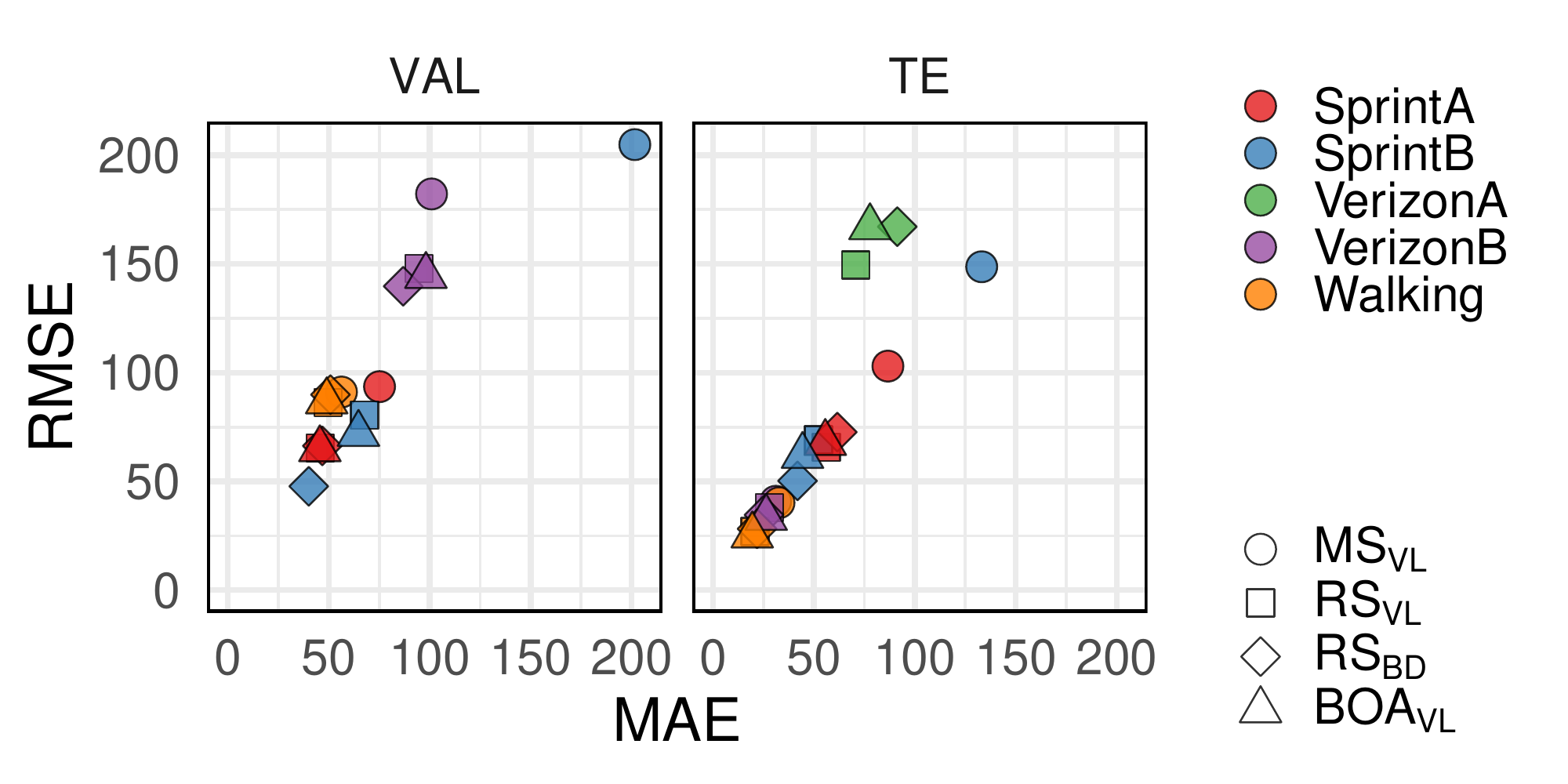} 
      \caption{\emph{5Gophers}. \ac{MAE} versus \ac{RMSE} colored by trace and grouped per configuration. Each subplot maps to the validation (\textit{VAL}) and testing (\textit{TE}) error, respectively.}
  \label{figure:5Gophers_error}
\end{figure*}

Figures \ref{fig:NYU-METS_error} and \ref{figure:5Gophers_error} portray the \ac{MAE} versus \ac{RMSE} for \emph{NYU-METS} and \emph{5Gophers}, respectively\footnote{For this work, we exclude T-Mobile since the reported measurements contain a multitude of invalid readings.}.
We use a diverging color palette and four point shapes to discriminate between the available traces configurations, respectively. 
Each individual point in the graph constitutes the model's \textit{operating point}. 
Since we study the performance from an error perspective, a good model operates on the bottom left area of the grid (and vice versa).
We report results for both validation (\textit{VAL}) and testing (\textit{TE}) sets.

\begin{figure*}[!htb]
\centering
  \begin{subfigure}[b]{.32\linewidth}
    \centering
    \includegraphics[width=.99\textwidth]{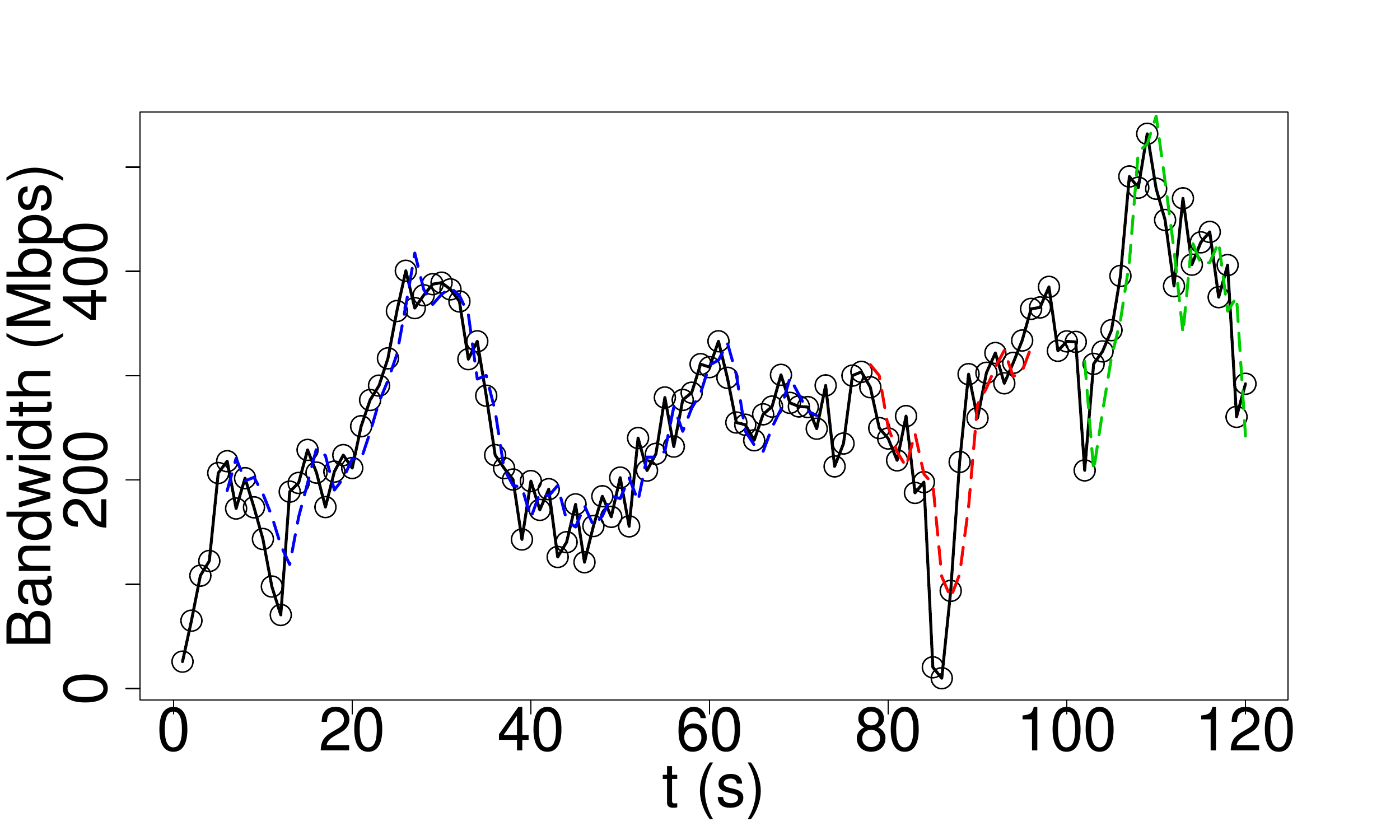}
    \caption{Sprint1}\label{fig:a2}
  \end{subfigure}%
  \begin{subfigure}[b]{.32\linewidth}
    \centering
    \includegraphics[width=.99\textwidth]{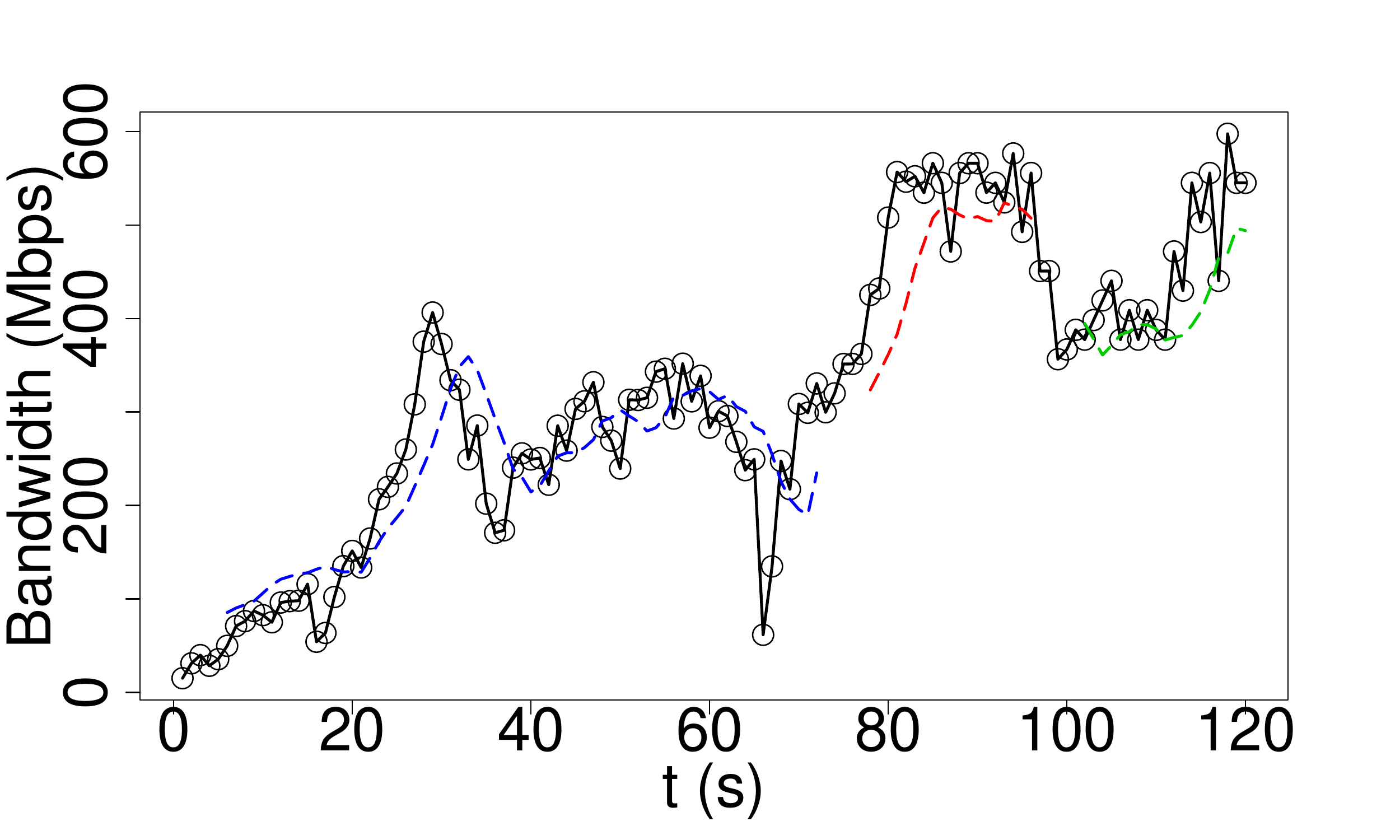}
    \caption{Sprint2}\label{fig:b2}
  \end{subfigure}%
  \begin{subfigure}[b]{.32\linewidth}
    \centering
    \includegraphics[width=.99\textwidth]{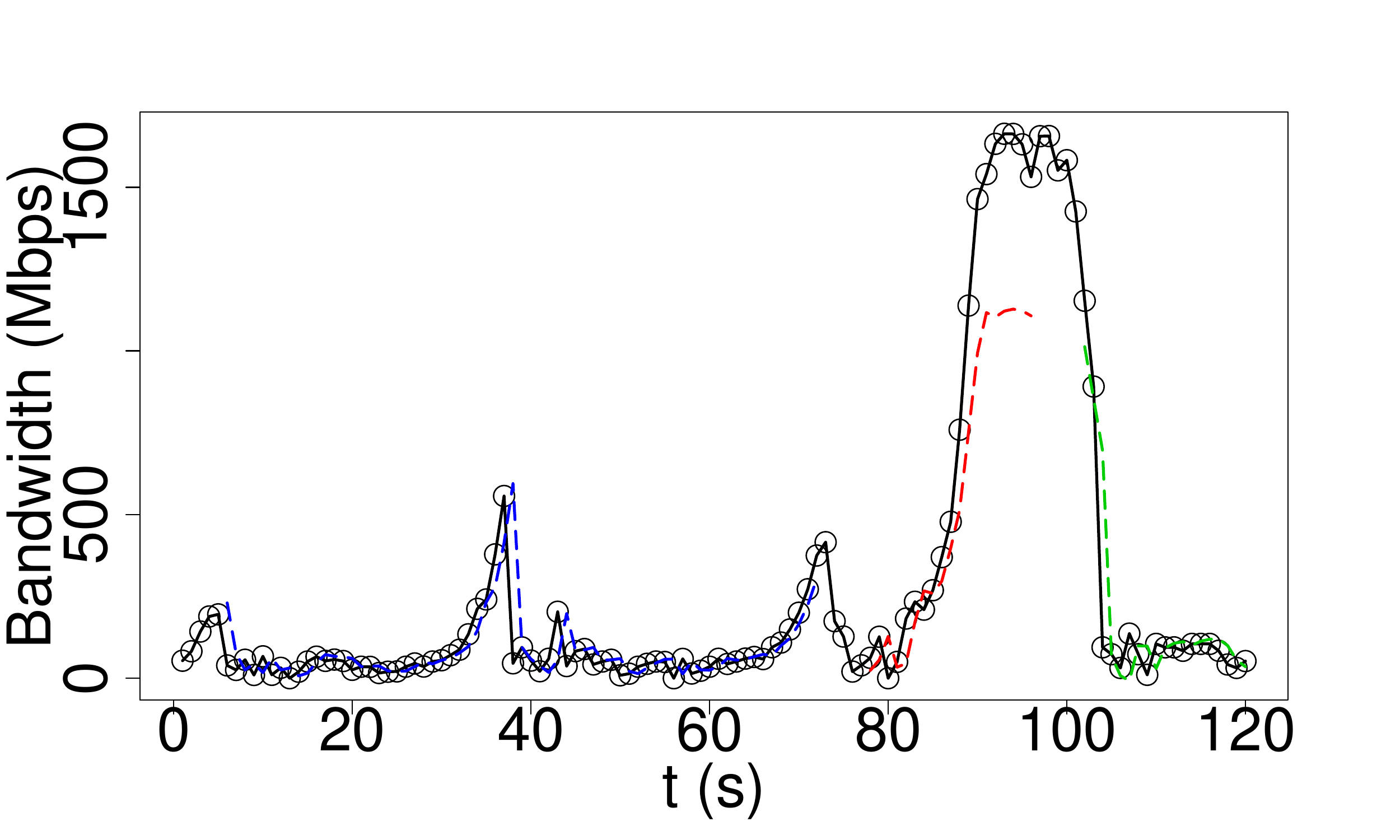}
    \caption{Verizon1}\label{fig:c2}
  \end{subfigure}
  \begin{subfigure}[b]{.32\linewidth}
    \centering
    \includegraphics[width=.99\textwidth]{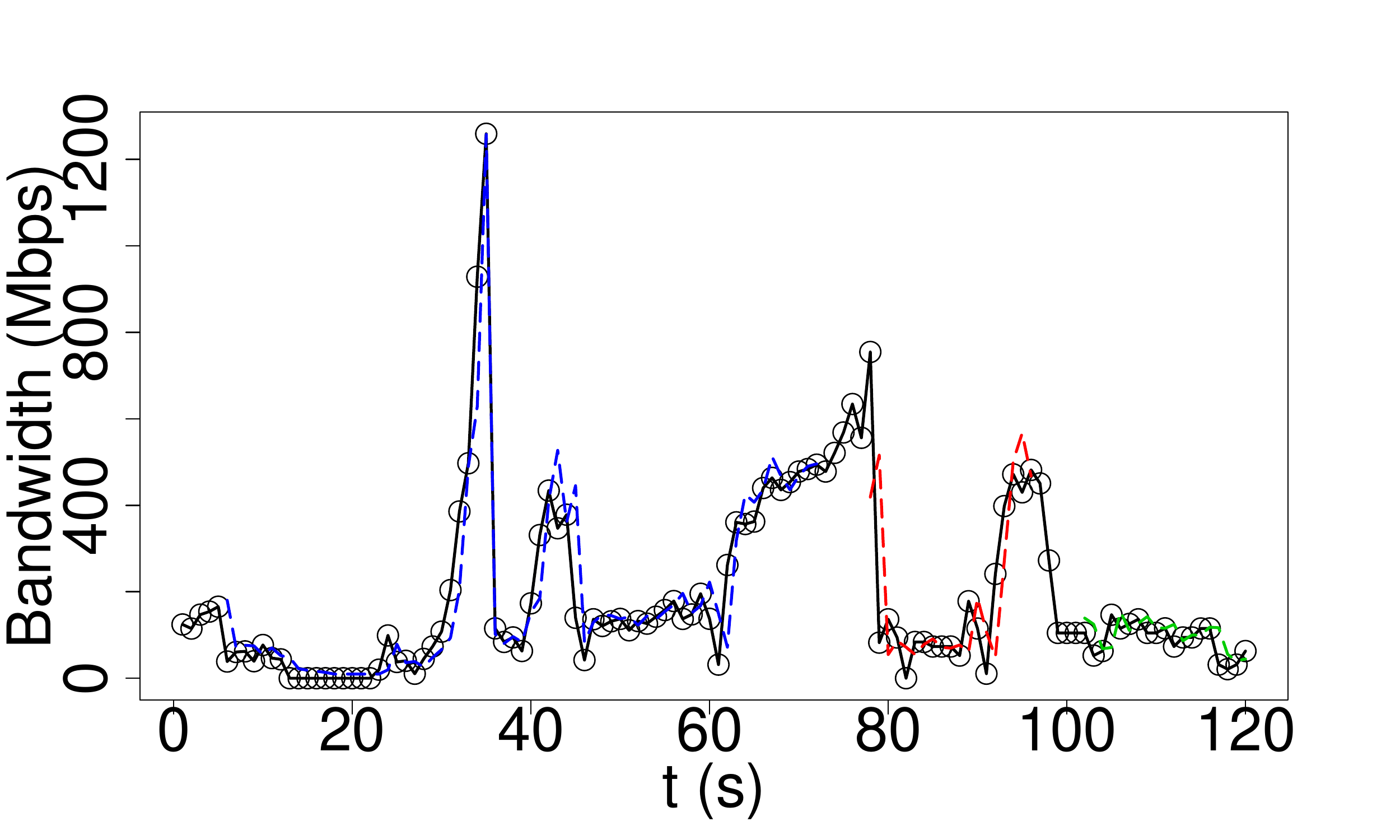}
    \caption{Verizon2}\label{fig:d2}
  \end{subfigure}%
  \begin{subfigure}[b]{.32\linewidth}
    \centering
    \includegraphics[width=.99\textwidth]{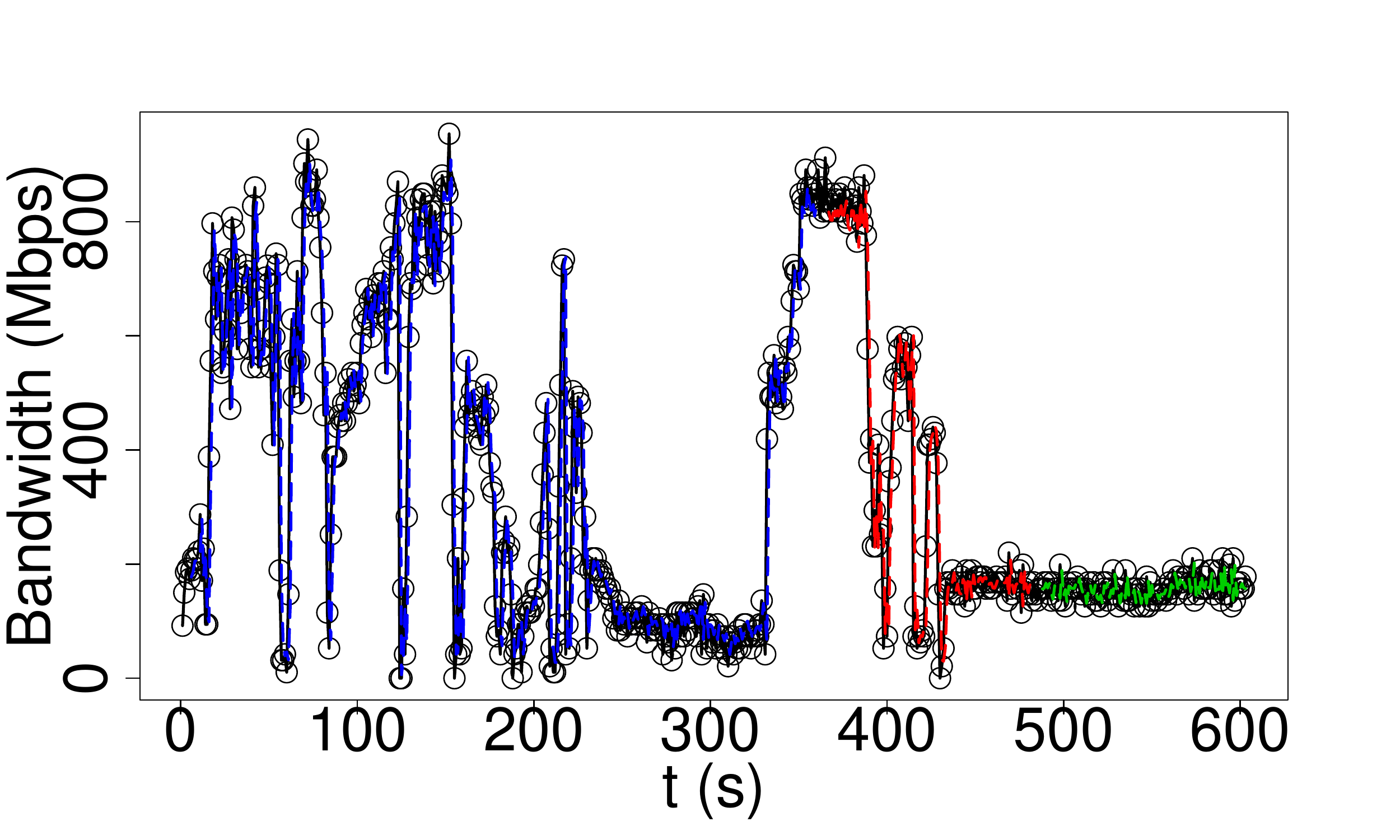}
    \caption{Walking trace}\label{fig:e2}
  \end{subfigure}%
  \caption{\emph{5Gophers} per-trace bandwidth measurements. \textit{y-axis} depicts the available bandwidth [Mbps] while \textit{x-axis} represents time [sec]. Groundtruth values are encoded in \textit{black}, while prediction values for  training, validation, and testing set, are colored as \textit{blue}, \textit{red}, and \textit{green}, respectively.}
  \label{fig:5Gophers}
\end{figure*}

At a glance, we observe that \emph{5Gophers} error show an $\approx100$-fold increase when compared to \emph{NUY-METS}.
This observation is in line with our expectations, considering the higher data rates that users experience in a \ac{5G} network environment (i.e., up to $1500$Mbps, see Figure \ref{fig:5Gophers}). 
Next, we follow up the results discussion per dataset. 

\textbf{NYU-METS}: From Figure \ref{fig:NYU-METS_error}, we make the following observations. 
The operating points of all models associated with the same trace are visually overlapping, hence, creating a clear-cut number of clusters.
This graph point distribution leads to a rather critical result.
In particular, it shows that different hyperparameter optimization algorithms have similar \ac{LSTM} performance across all \ac{4G} traces.
We verify the significance difference between the models using Dunn's non parametric pairwise test which outputs $p-values>0.95$.
We then quantify how much improvement hyperparameter optimization can bring compared to the selected hyperparameter $MS_{VL}$ which is optimized for 4G traces \cite{mei2019realtime}. We observe testing \ac{MAE} improvement up to $6.46\%$ with an average of $\approx3.5\%$.

\begin{figure*}[!htb]
\centering
  \begin{subfigure}[b]{.32\linewidth}
    \centering
    \includegraphics[width=.99\textwidth]{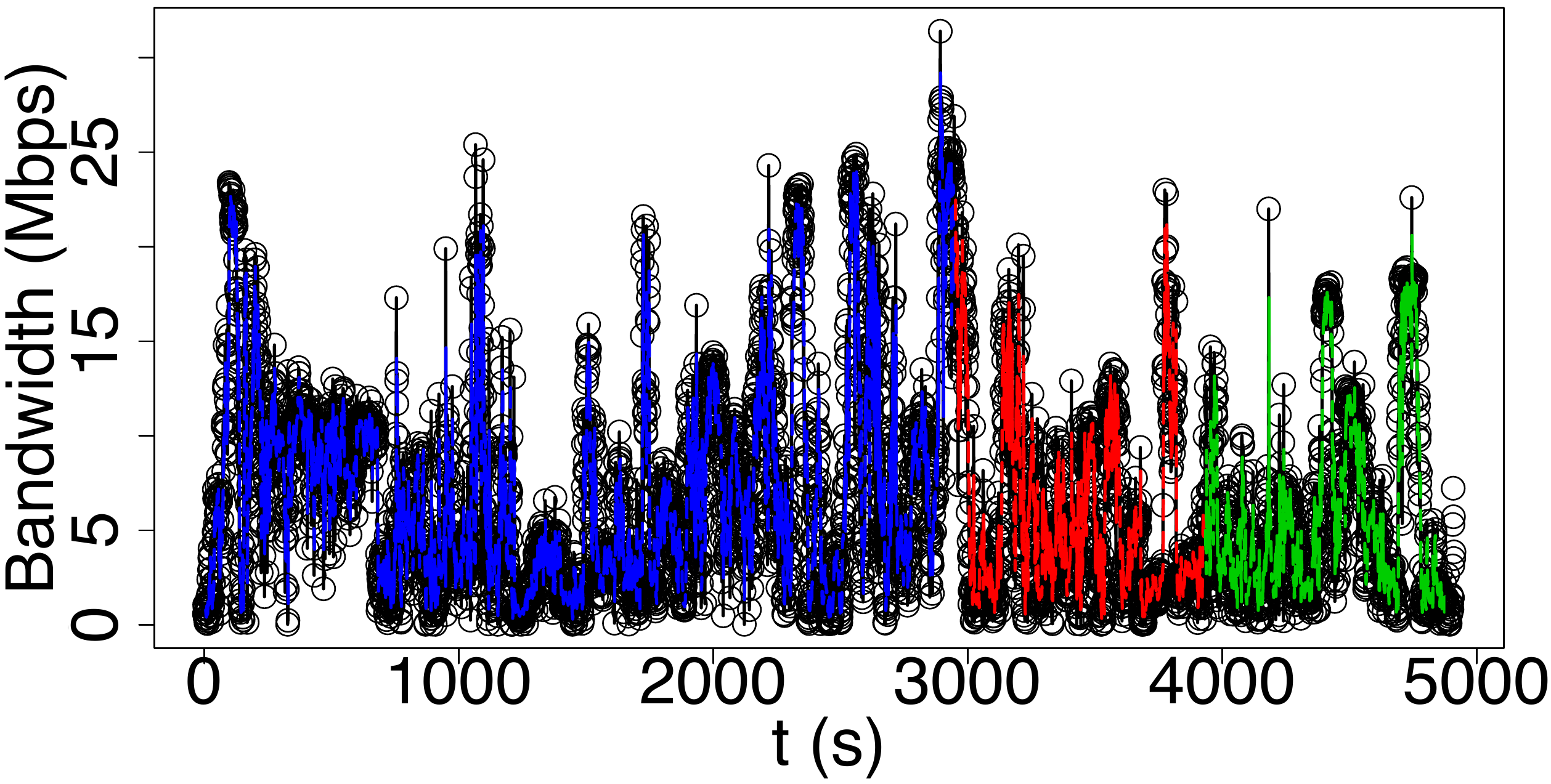}
    \caption{7TrainA}\label{fig:a1}
  \end{subfigure}%
  \begin{subfigure}[b]{.32\linewidth}
    \centering
    \includegraphics[width=.99\textwidth]{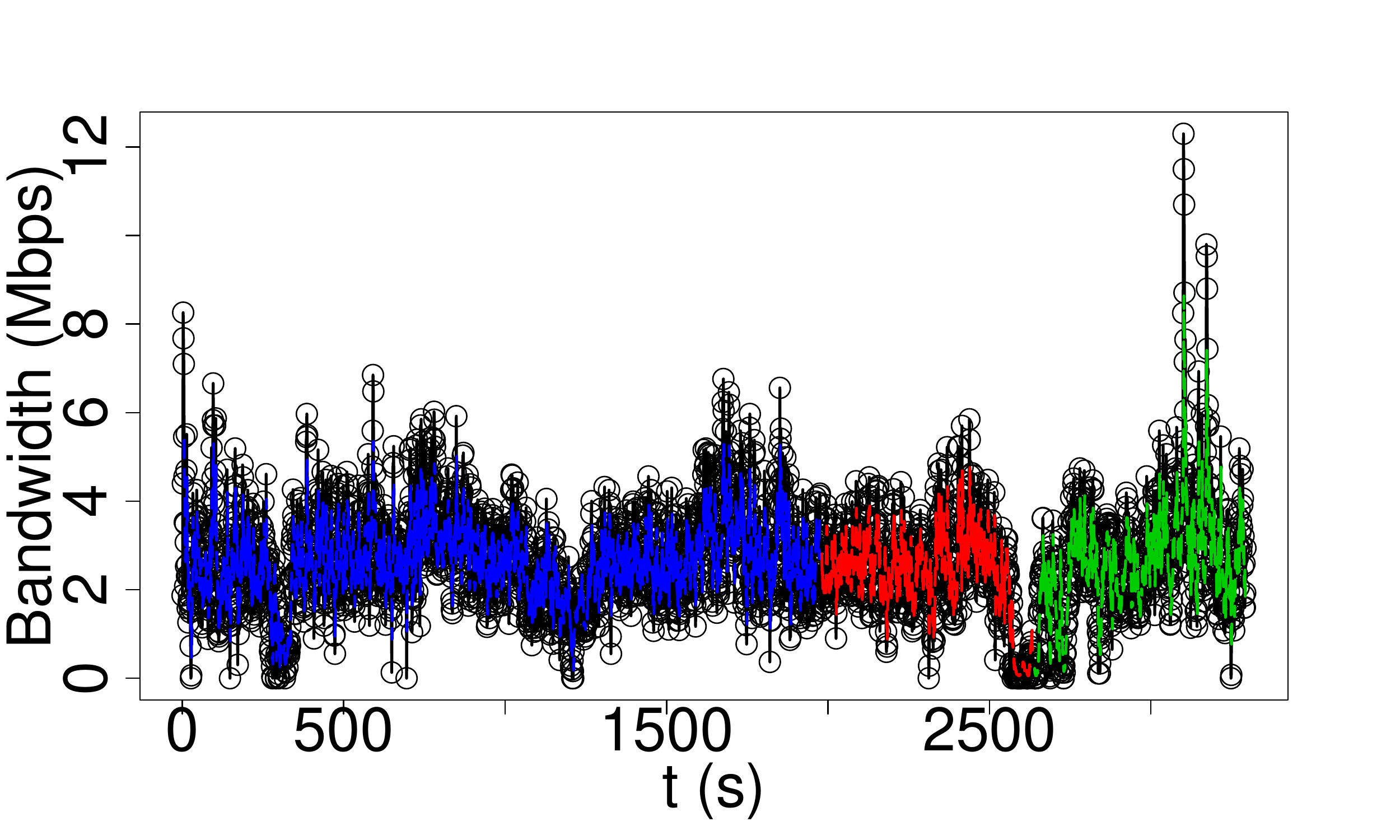}
    \caption{Bus62A}\label{fig:d1}
  \end{subfigure}%
  \begin{subfigure}[b]{.32\linewidth}
    \centering
    \includegraphics[width=.99\textwidth]{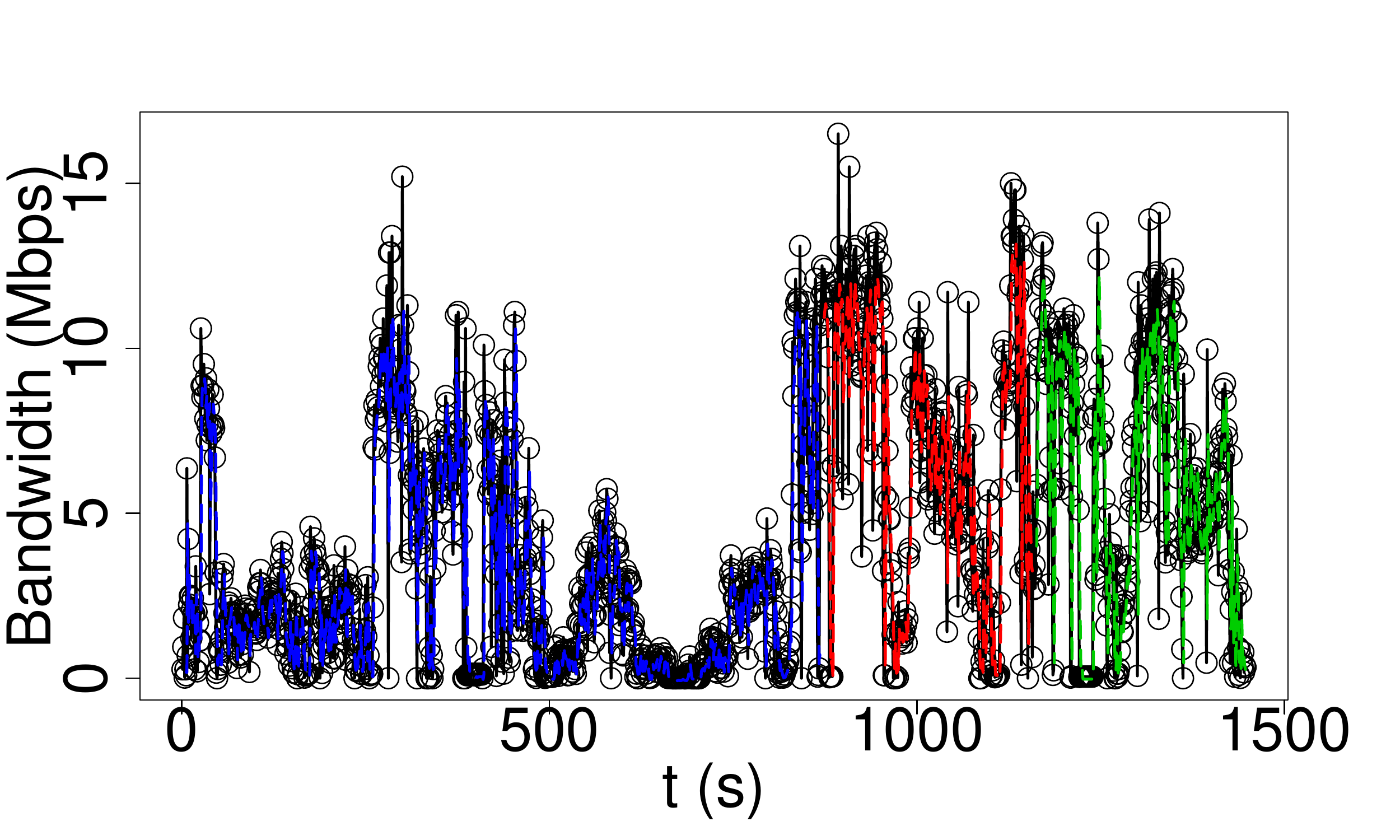}
    \caption{LI RR}\label{fig:g1}
  \end{subfigure}%
  \caption{\emph{NYU-METS} per-trace bandwidth measurements. \textit{y-axis} depicts the available bandwidth [Mbps] while \textit{x-axis} represents time [sec]. Groundtruth values are encoded in \textit{black}, while prediction values for  training, validation, and testing set, are colored as \textit{blue}, \textit{red}, and \textit{green}, respectively.}\label{fig:NYU-METS}
\end{figure*}

Furthermore, we observe that \ac{LSTM} models operate on a low \ac{MAE} region which is the result of the low-variance data rates experienced in \ac{4G} under mobility, thus restricting the room for vast performance improvement.
Across the two dataset portions, we find that, in average, validation error is lower than testing error, which is the common scenario, since model tuning takes place in the former set. 
Last, Figure \ref{fig:NYU-METS} illustrates the bandwidth time series graph of three \emph{NYU-METS} traces\footnote{We do not report all seven traces due to space considerations.}.
Color is used for mapping the forecasting lines ( ($RS_{VL}$)) with the associated part of the trace, i.e., training, validation, and testing set.
As evident, \ac{LSTM} models follow the data trends along the traces with minimal error deviation. 

\textbf{5Gophers}: As evident from Figure \ref{figure:5Gophers_error}, \ac{LSTM} models do not maintain the same clustering properties as in Figure \ref{fig:NYU-METS_error}, instead, they are higher dispersed across the graph.
In particular, we observe that for each mobility trace, all $MS_{VL}$ configurations operate in a higher error region (i.e., upper right).
In this regard, the average testing \ac{MAE} improvement between $MS_{VL}$ and the remainder \ac{LSTM} configurations for Sprint and Verizon is $49\%$ ($33\%$ for SprintA and $65\%$ for SprintB) and $26.5\%$ ($38\%$ for VerizonA and $15\%$ for VerizonB), respectively.
In addition, the testing \ac{MAE} improvement for the walking trace is also substantial ($\approx37\%$), which highlights the need for hyperparameter optimization even in lower mobility scenarios. 
These results clearly show that 5G has very different characteristics compared to 4G and the set of hyperparameters optimized for 4G cannot be directly used in 5G settings.

Furthermore, we compare the average performance between the two hyperparameter optimization algorithms.
To provide a fair comparison and remove any undesired biases, we only account for $RS_{VL}$ and $BOA_{VL}$.
We find that across all traces, including the walking trace, \ac{BOA} provides a slight performance boost of $\approx4\%$.

Next, we study the performance of Bidirectional \ac{LSTM} networks to infer whether they bring any significant benefits to the bandwidth forecasting problem. 
Likewise, we compare $RS_{VL}$ and $RS_{BD}$ to remove the hyperparameter optimization bias. 
On the one hand, we find that Bidirectional \ac{LSTM} networks outperform the Vanilla counterpart in two out of the four driving traces (i.e., SprintA and VerizonA) with an average testing \ac{MAE} improvement of $\approx15\%$.
On the other hand, however, they seem to be performing  worse in the other two driving traces (i.e., SprintB and VerizonB) by an average testing \ac{MAE} percentage of $\approx15\%$.
This finding verifies the fact that  bidirectional \ac{LSTM} networks are very dependant on the respective dataset underlying characteristics.
Therefore, the question of whether or not they are fitting for a given application can only be answered after systematic empirical analysis.

Tables \ref{table:5Gophers} and \ref{table:NYU-METS} provide additional insights with respect to the \ac{RT} complexity. 
In particular, each row reports the \ac{RT} [min] per trace and across the available configurations.
In average, we observe that $RS_{BD}$ present a higher \ac{RT} of $\approx54\%$ when compared to the $RS_{VL}$.
The above result is not balanced across the two datasets due to the randomness bias during the hyperparameter selection process. 
For example, instances of the algorithm where higher values for \acp{HL} or neurons are selected will require more computational power to accomplish. 
On the other hand, the average increase in \ac{RT} between $BOA_{VL}$ and $RS_{VL}$ reaches to a percentage of $\approx154\%$, which can be attributed to the following.
First, since \ac{BOA} leverages the Bayes Theorem for minimizing the gradient descent error, it requires more time to reach to a good solution, opposed to \ac{RS} which randomly scans the search space. 
And second, the \textit{R} \ac{BOA} implementation is not optimized to run on top of a GPU, which adds to the \ac{RT} overhead.

\textbf{Takeaways:} The hyperparameter optimization paradigm is critical for improving several performance aspects in \ac{LSTM} networks. 
However, its significance is highly dependant on the underlying data attributes.
For example, sequences that display large signs of variance over time are notably harder to predict, thus requiring hyperparameter configurations of high precision.
On the contrary, sequences that either meet the stationarity test requirements or show consistent trends across time have limited gains, since most combinations of hyperaparameters within the predefined search range will provide comparable performance. 

For the bandwidth prediction task under study, therefore, we find that hyperparameter optimization is especially vital in \ac{5G} networks, since they introduce higher data rates and larger variations over time. 
Coupled with the mobility aspect, a suboptimal hyperparameter set will fail to capture the intrinsic network characteristics. Among the two available optimization solutions under study, we observe that \ac{BOA} provides slightly better results, since it utilizes probability theory concepts for minimizing the error. 
We further observe that \ac{4G} networks offer significantly lower data rates with a lower variation factor which significantly diminishes the need for advanced hyperparameter optimization solutions.

\subsection{Impact of hyperparameters} 

\begin{figure*}[!htb]
  \centering
    \includegraphics[keepaspectratio,width = 0.85\linewidth]{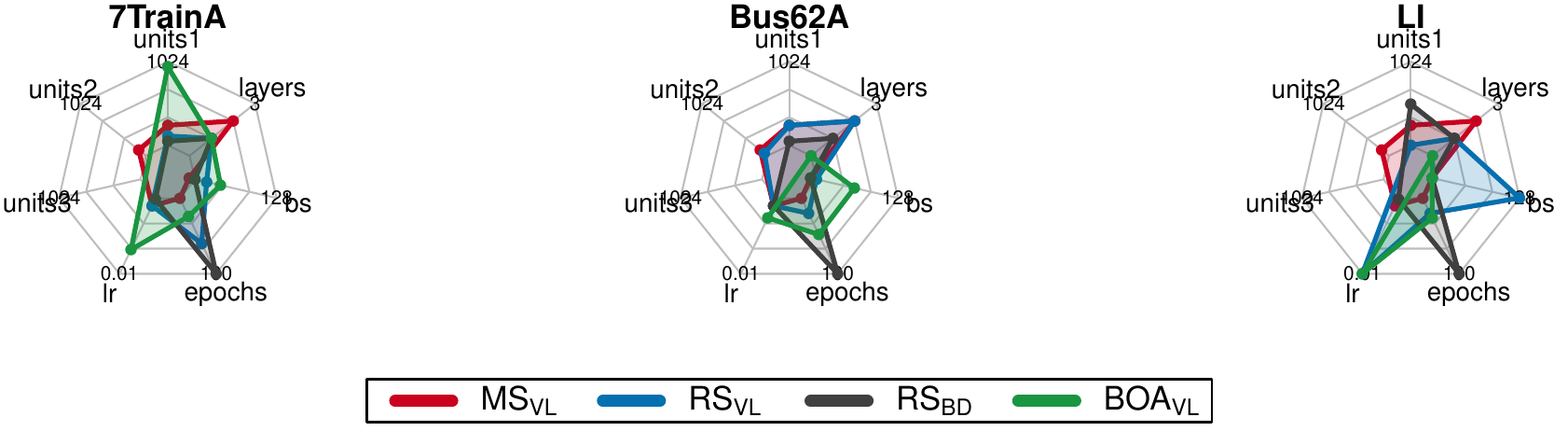} 
      \caption{\emph{NYU-METS}. Each radar plot shows the respective trace's hyperparameter values. Color is used to discriminate between the configurations under test.}
  \label{fig:NYU-METS_radar}
\end{figure*}

\begin{figure*}[!htb]
  \centering
    \includegraphics[keepaspectratio,width = 0.95\linewidth]{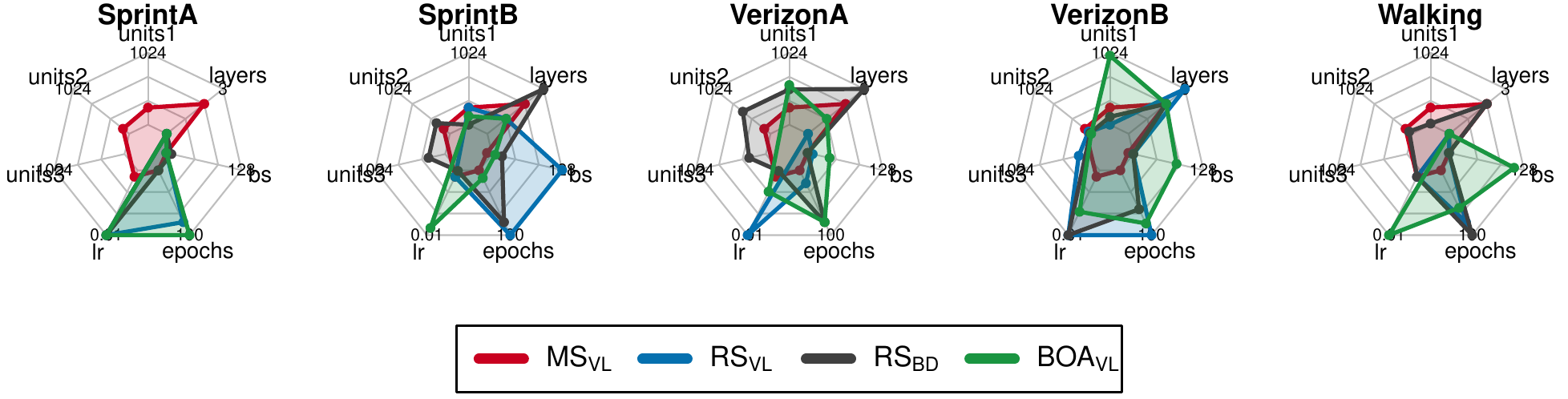}
      \caption{\emph{5Gophers}. Each radar plot shows the respective hyperparameter values for each trace. Color is used to discriminate between the configurations under test.}
  \label{figure:5gophers_radar}
\end{figure*}

The current subsection discusses the hyperparameter selection process and investigates potential trends or patterns across the available traces and datasets. 
Figures \ref{fig:NYU-METS_radar} and \ref{figure:5gophers_radar} illustrate the selected hyperparameters along the selected \emph{NYU-METS} traces 
and \emph{5Gophers} datasets, respectively.
The edges of each radar plot map to the maximum search range values across the seven hyperparameters, while color is used to discriminate between each configuration. 

Overall, it is evident that the selected \ac{LSTM} models adopt disparate strategies for minimizing the gradient descent error function, a result that verifies the high degree of complexity in the neural network ecosystem. 
Next, we isolate a number of use cases per dataset in an effort to discover any hidden patterns or trends alongside the pool of hyperparameters.

\textbf{NYU-METS:} One of the most notable remarks across all available \ac{4G} traces is that none of the \ac{LSTM} models utilize more than two \acp{HL}.
This observation infers than the addition of a third \ac{HL} would add excessive noise to the learning model rather than minimizing the predictive error.
The average number of neurons per \ac{HL} across all traces equals to $258$ and $81$ for \ac{HL}-1 and \ac{HL}-2, respectively.
In addition, more than half of the models feature a \ac{LR} that rests around the $0.001$ region with an average value of $0.002$, the average \ac{BS} is $43$, while the average number of epochs equals to $65$, revealing a relatively early error saturation point.  
The above numbers that \ac{LSTM} networks opt to stray away from extreme hyperparameter values in an attempt to regulate the variance bias tradeoff. 

\textbf{5Gophers:} Exploring potential patterns across the \ac{5G} bandwidth measurements, we observe that in some scenarios the optimized \ac{LSTM} models feature a third \ac{HL}. 
Again, this is a rather expected result considering the higher data rates and variation in \ac{5G} in addition with the challenging bandwidth trends over time and under mobility. 
Likewise, the average number of neurons per \ac{HL} across all layers equals to $197$, $79$, and $32$ for \ac{HL}-1, \ac{HL}-2 and \ac{HL}-3, respectively.
We observe that the number of neurons per \ac{HL} for both datasets follow a decreasing trend across the network, which also coincides with the neuron selection from authors in \cite{mei2019realtime} (i.e., HL-1=$256$ and HL-2=$128$). 
Additionally, the average \ac{LR} and \ac{BS} across all traces equal to $0.004$ and $28$, respectively, while the number of epochs averages $76$, which is an $\approx17\%$ increase comparing to the \ac{4G} counterpart result hinting a slower gradient descent convergence.

\textbf{Takeaways:} The designated results reveal that hyperparameter optimization is crucial for achieving superior performance in \ac{5G} networks, while trading computational resources.
The above finding is further reinforced by the reported model hyperparameters, which significantly vary across technology standards, mobility scenarios, and to some extent network operators.
Across the two datasets, we find that \ac{5G} data require \ac{LSTM} configurations with an increased number of \acp{HL} and epochs, while featuring significantly lower \ac{BS} values, which further reveals the challenging network conditions in such scenarios.

\subsection{Challenges and Limitations} 

We further report a list of limitations and challenges that should be considered moving forward. 
First, the number of iterations for both hyperparameter algorithms is set to a constant value, i.e., $50$ in both cases. 
This decision was made after closely observing the fast saturation rate of the gradient descent error.
However, we argue that future \ac{5G} applications may require a higher number of iterations to converge to a good solution, although, this would result to a substantial increase of the \ac{RT} complexity. 
Second, the hyperparameter search range can be stretched to accommodate a higher pool of values, which could potentially add to the performance.
The selected values were decided as a compromise to the performance versus \ac{RT} complexity tradeoff.
Last, both datasets are composed of traces with a limited number of samples 
which can have a negative effect on the \ac{LSTM} performance since they are known to perform best with larger datasets. 

\section{Conclusions and Future Work}
\label{sec:conclusions}

In this paper, we studied the challenging task of bandwidth forecasting in next-generation \ac{MBB} networks under mobility.
To ease our goal, we designed \emph{HINDSIGHT++}, an open-source \textit{R}-based framework that allows for \ac{LSTM} experimentation in time series data.
We specifically focused on the hyperparameter optimization aspect which is critical for achieving state-of the performance.
We analysed and performed a comparative analysis between two open-source datasets of bandwidth measurements in operational \ac{4G} and \ac{5G} networks, respectively, aiming to quantify the \ac{LSTM} performance improvement under different configuration setups.
Results show that hyperparameter optimization provides significant benefits under \ac{5G} settings compared to \ac{4G} settings and the optimal parameters for \ac{4G} cannot be directly applied considering the substantially higher data rates and variation over time that users experience in \ac{5G} network conditions. 
As for the future, we plan on integrating feature selection algorithms, additional \ac{LSTM} variants (e.g., \ac{CNN} \ac{LSTM} and multiplicative \ac{LSTM} \cite{krause2016multiplicative}), and provide support for supervised classification tasks.
Furthermore, we will carry out a dedicated measurement campaign to collect additional network features (e.g., signal strength, latency, etc.). Using the new data, we aim to explore potential correlation or causation relationships and show whether or not they bring any significant gains to the \ac{LSTM} performance.

\section*{Acknowledgments}
\label{acknowledgments}

This work is partially funded by the EU H2020 5GENESIS (815178), and by the Norwegian Research Council project MEMBRANE (250679).

\newpage
\thispagestyle{empty}

\begin{wrapfigure}{l}{25mm} 
\includegraphics[width=1in,height=1.5in,clip,keepaspectratio]{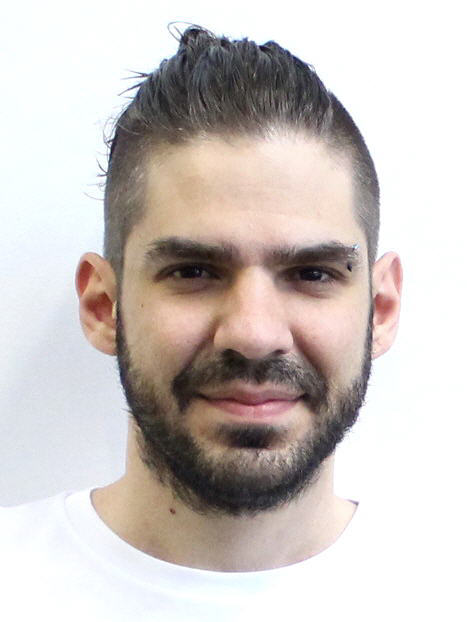}
\end{wrapfigure} 
\textbf{Konstantinos Kousias} is a Ph.D. Student at the Faculty of Mathematics and Natural Sciences in University of Oslo affiliated with Simula Research Laboratory. 
He received his B.Sc. and M.Sc. degrees from the Department of Electrical and Computer Engineering at University of Thessaly in Volos, Greece. 
His research focuses on the empirical modeling and evaluation of mobile networks and Internet of Things (IoT) using AI driven analytics.
\newline

\begin{wrapfigure}{l}{25mm} 
\includegraphics[width=1in,height=1.5in,clip,keepaspectratio]{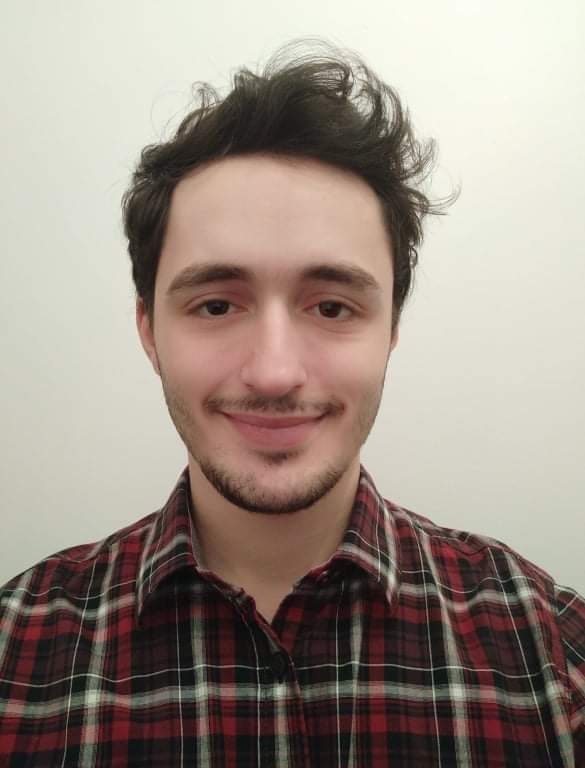}
\end{wrapfigure} 
\textbf{Apostolos Pappas} received his Bachelor diploma from the Department of Electrical and Computer Engineering at University of Thessaly, Volos, Greece in 2020.
His research interests lie in Communication Systems and Machine Learning.
During the summer of 2019, he worked as an intern at Simula Research Laboratory, mainly focusing on the development of LSTM networks for time series applications on mobile networks and expanding the \textit{Hindsight} framework.
\newline

\begin{wrapfigure}{l}{25mm} 
\includegraphics[width=1in,height=1.5in,clip,keepaspectratio]{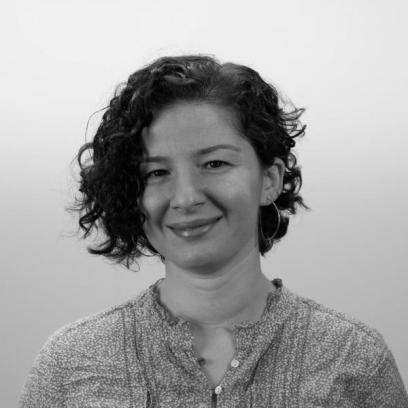}
\end{wrapfigure} 
\textbf{\"Ozg\"u Alay} received the B.S. and M.S. degrees in Electrical and Electronic Engineering from Middle East Technical University, Turkey, and Ph.D. degree in Electrical and Computer Engineering at Tandon School of Engineering at New York University. Currently, she is an Associate Professor in University of Oslo, Norway and Head of Department at Mobile Systems and Analytics (MOSAIC) of Simula Metropolitan, Norway. Her research interests lie in the areas of mobile broadband networks including 5G, low latency networking, multipath protocols and robust multimedia transmission over wireless networks. She is author of more than 70 peer-reviewed IEEE and ACM publications and she actively serves on technical boards of major conferences and journals.
\newline

\begin{wrapfigure}{l}{25mm} 
\includegraphics[width=1in,height=1.5in,clip,keepaspectratio]{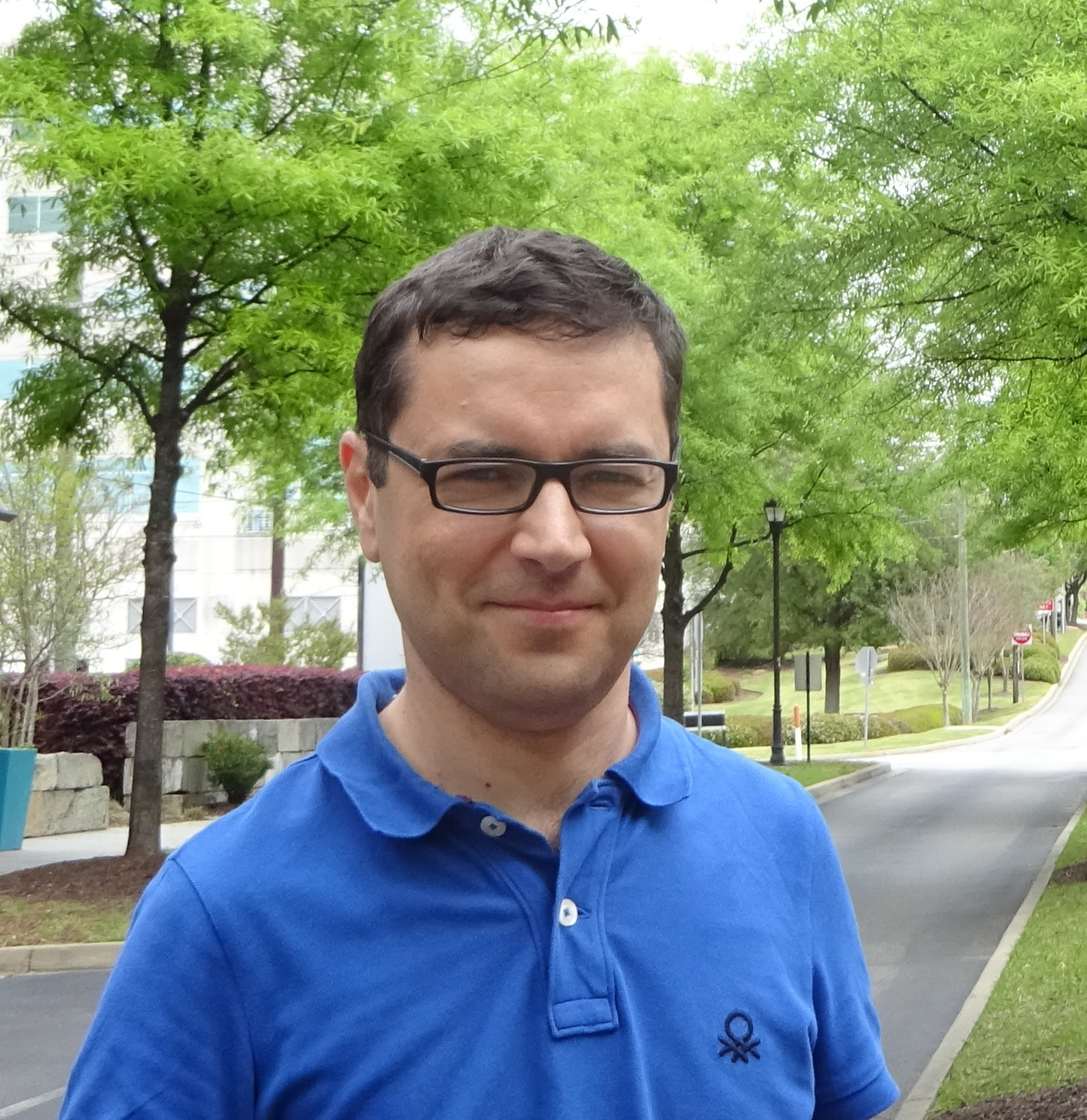}
\end{wrapfigure} 
\textbf{Antonios Argyriou} received the Diploma in electrical and computer engineering from Democritus University of Thrace, Greece, and the M.S. and Ph.D. degrees in electrical and computer engineering as a Fulbright scholar from the Georgia Institute of Technology, Atlanta, USA. Currently, he is an Associate Professor at the department of electrical and computer engineering, University of Thessaly, Greece. From 2007 until 2010 he was a Senior Research Scientist at Philips Research, Eindhoven, The Netherlands. Dr. Argyriou serves in the TPC of several international conferences and workshops. His research interests are in the areas of communication systems, and statistical signal processing.
\newline

\begin{wrapfigure}{l}{25mm} 
\includegraphics[width=1in,height=1.5in,clip,keepaspectratio]{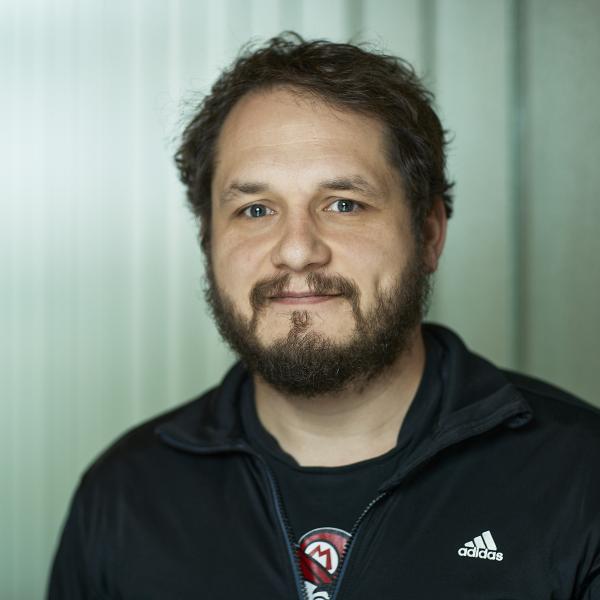}
\end{wrapfigure} 
\textbf{Michael Alexander Riegler}, Chief Research Scientist, SimulaMet, holds degrees in computer science from the University of Oslo (UiO) and Klagenfurt University. He is experienced in medical multimedia data analysis and understanding, image processing, image retrieval, parallel processing, crowdsourcing, social computing and user intent. He is a recognized expert in his field, serving as a member of the Norwegian Council of Technology on Machine Learning for Healthcare.
\newline

\end{document}